\documentclass[twoside,twocolumn,9pt]{article}
\usepackage{extsizes}
\usepackage[super,sort&compress,comma]{natbib} 
\usepackage[version=3]{mhchem}
\usepackage[left=1.5cm, right=1.5cm, top=1.785cm, bottom=2.0cm]{geometry}
\usepackage{balance}
\usepackage{widetext}
\usepackage{times,mathptmx}
\usepackage{sectsty}
\usepackage{graphicx} 
\usepackage{lastpage}
\usepackage[format=plain,justification=raggedright,singlelinecheck=false,font={stretch=1.125,small,sf},labelfont=bf,labelsep=space]{caption}
\usepackage{float}
\usepackage{fancyhdr}
\usepackage{fnpos}
\usepackage[english]{babel}
\usepackage{array}
\usepackage{droidsans}
\usepackage{charter}
\usepackage[T1]{fontenc}
\usepackage[usenames,dvipsnames]{xcolor}
\usepackage{setspace}
\usepackage[compact]{titlesec}
\usepackage{textcomp}
\usepackage{multirow}

\usepackage{epstopdf}

\definecolor{cream}{RGB}{222,217,201}

\begin{document}

\pagestyle{fancy}
\thispagestyle{plain}
\fancypagestyle{plain}{

\fancyhead[C]{\includegraphics[width=18.5cm]{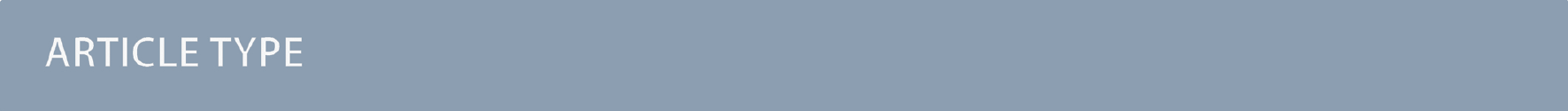}}
\fancyhead[L]{\hspace{0cm}\vspace{1.5cm}\includegraphics[height=30pt]{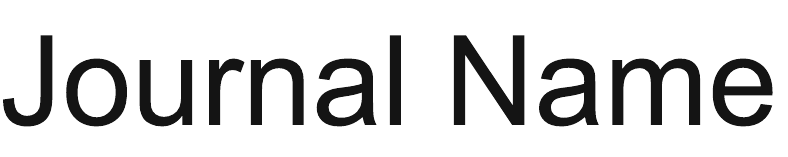}}
\fancyhead[R]{\hspace{0cm}\vspace{1.7cm}\includegraphics[height=55pt]{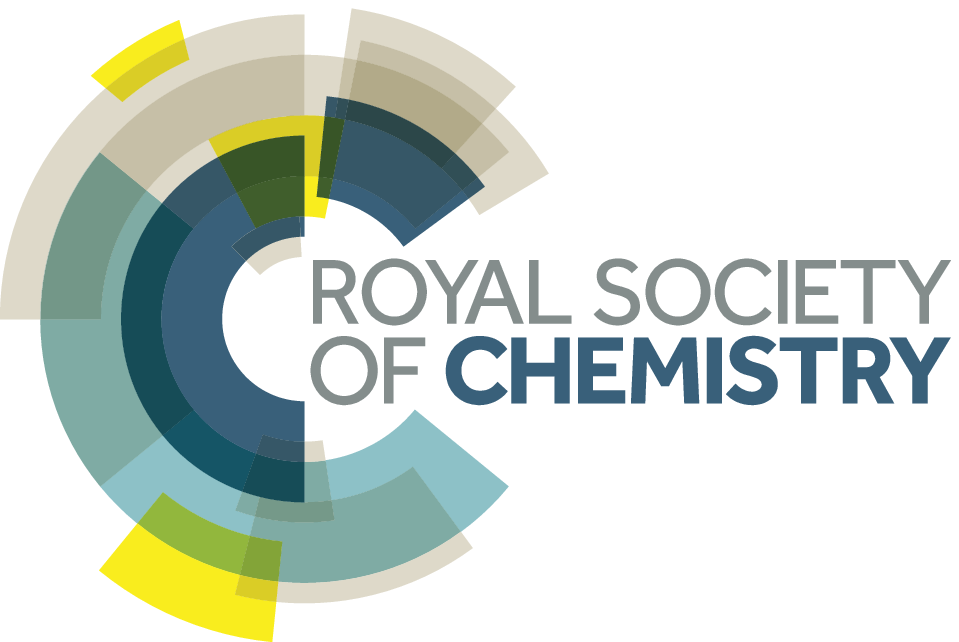}}
\renewcommand{\headrulewidth}{0pt}
}

\makeFNbottom
\makeatletter
\renewcommand\LARGE{\@setfontsize\LARGE{15pt}{17}}
\renewcommand\Large{\@setfontsize\Large{12pt}{14}}
\renewcommand\large{\@setfontsize\large{10pt}{12}}
\renewcommand\footnotesize{\@setfontsize\footnotesize{7pt}{10}}
\makeatother

\renewcommand{\thefootnote}{\fnsymbol{footnote}}
\renewcommand\footnoterule{\vspace*{1pt}%
\color{cream}\hrule width 3.5in height 0.4pt \color{black}\vspace*{5pt}} 
\setcounter{secnumdepth}{5}

\makeatletter 
\renewcommand\@biblabel[1]{#1}            
\renewcommand\@makefntext[1]%
{\noindent\makebox[0pt][r]{\@thefnmark\,}#1}
\makeatother 
\renewcommand{\figurename}{\small{Fig.}~}
\sectionfont{\sffamily\Large}
\subsectionfont{\normalsize}
\subsubsectionfont{\bf}
\setstretch{1.125} 
\setlength{\skip\footins}{0.8cm}
\setlength{\footnotesep}{0.25cm}
\setlength{\jot}{10pt}
\titlespacing*{\section}{0pt}{4pt}{4pt}
\titlespacing*{\subsection}{0pt}{15pt}{1pt}

\fancyfoot{}
\fancyfoot[LO,RE]{\vspace{-7.1pt}\includegraphics[height=9pt]{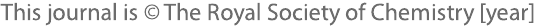}}
\fancyfoot[CO]{\vspace{-7.1pt}\hspace{13.2cm}\includegraphics{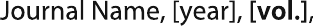}}
\fancyfoot[CE]{\vspace{-7.2pt}\hspace{-14.2cm}\includegraphics{head_foot/RF}}
\fancyfoot[RO]{\footnotesize{\sffamily{1--\pageref{LastPage} ~\textbar  \hspace{2pt}\thepage}}}
\fancyfoot[LE]{\footnotesize{\sffamily{\thepage~\textbar\hspace{3.45cm} 1--\pageref{LastPage}}}}
\fancyhead{}
\renewcommand{\headrulewidth}{0pt} 
\renewcommand{\footrulewidth}{0pt}
\setlength{\arrayrulewidth}{1pt}
\setlength{\columnsep}{6.5mm}
\setlength\bibsep{1pt}

\makeatletter 
\newlength{\figrulesep} 
\setlength{\figrulesep}{0.5\textfloatsep} 

\newcommand{\topfigrule}{\vspace*{-1pt}%
\noindent{\color{cream}\rule[-\figrulesep]{\columnwidth}{1.5pt}} }

\newcommand{\botfigrule}{\vspace*{-2pt}%
\noindent{\color{cream}\rule[\figrulesep]{\columnwidth}{1.5pt}} }

\newcommand{\dblfigrule}{\vspace*{-1pt}%
\noindent{\color{cream}\rule[-\figrulesep]{\textwidth}{1.5pt}} }

\makeatother

\twocolumn[
  \begin{@twocolumnfalse}
\vspace{3cm}
\sffamily
\begin{tabular}{m{4.5cm} p{13.5cm} }

\includegraphics{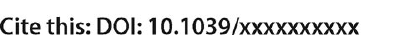} & \noindent\LARGE{\textbf{Ion-Induced Molecular Growth in Clusters of Small \mbox{Hydrocarbon} Chains}} \\
\vspace{0.3cm} & \vspace{0.3cm} \\

 & \noindent\large{Michael Gatchell,$^{\ast}$\textit{$^{a}$} Rudy Delaunay,\textit{$^{b}$} Giovanna D'Angelo,\textit{$^{a,c,d}$} Arkadiusz Mika,\textit{$^{b}$} \mbox{Kostiantyn} Kulyk,\textit{$^{a}$} Alicja Domaracka,\textit{$^{b}$} Patrick Rousseau,\textit{$^{b}$} Henning Zettergren,\textit{$^{a}$} Bernd A.\ Huber,\textit{$^{b}$} and Henrik Cederquist\textit{$^{a}$}} \\

\includegraphics{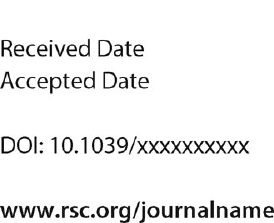} & \noindent\normalsize{We report on studies of collisions between 3\,keV Ar$^+$ projectile ions and neutral targets of isolated 1,3-butadiene (C$_4$H$_6$) molecules and cold, loosely bound clusters of these molecules. We identify molecular growth processes within the molecular clusters that appears to be driven by knockout processes and that could result in the formation of (aromatic) ring structures. These types of reactions are not unique to specific projectile ions and target molecules, but will occur whenever atoms or ions with suitable masses and kinetic energies collide with aggregates of matter, such as carbonaceous grains in the interstellar medium or aerosol nanoparticles in the atmosphere. } \\

\end{tabular}

 \end{@twocolumnfalse} \vspace{0.6cm}

  ]

\renewcommand*\rmdefault{bch}\normalfont\upshape
\rmfamily
\section*{}
\vspace{-1cm}


\footnotetext{\textit{$^{a}$~Department of Physics, Stockholm University, SE-106 91 Stockholm, Sweden; E-mail: gatchell@fysik.su.se}}
\footnotetext{\textit{$^{b}$~Normandie Univ., ENSICAEN, UNICAEN, CEA, CNRS, CIMAP, 14000 Caen, France}}
\footnotetext{\textit{$^{c}$~UCIBIO, REQUIMTE, Departamento de Qu{\'i}mica e Bioqu{\'i}mica, Universidade do Porto, Porto, Portugal}}
\footnotetext{\textit{$^{d}$~Universidad Aut{\'o}noma de Madrid, Departamento de Qu{\'i}mica, 28049 Madrid, Spain}}




\section{Introduction}
Carbon forms the basis of the majority of the molecular species that so far have been identified in space\cite{Tielens:2013aa}. Although small carbon-based molecules, like CO, are some of the most abundant molecules in space, only a small fraction of the carbon is expected to be locked up in such species\cite{Tielens:2013aa}. Instead a large portion of the interstellar carbon---perhaps more than 10 percent---is expected to be contained in some of the largest molecules identified in space, such as the Polycyclic Aromatic Hydrocarbons (PAHs) and the fullerenes\cite{IAU:8465112,Berne:2012aa}. Since the 1980s\cite{1984A&A...137L...5L,1989ApJS...71..733A,Tielens:2008aa}, PAH molecules have been suggested to be carriers of unidentified emission bands in the near infrared that are observed on top of the IR continua from dust grains in many regions of space. These bands are generally considered to arise from the characteristic C--C and C--H stretching modes of aromatic rings, but because these signatures are nearly the same for all PAH types, no individual PAH species has been identified so far in the interstellar medium (ISM)\cite{Tielens:2013aa}. Individual fullerenes (C$_{60}$ and C$_{70}$), on the other hand, have been identified in the IR emission spectra of planetary nebulae\cite{Cami:2010aa} and reflection nebulae\cite{Sellgren:2010aa}, and more recently C$_{60}^+$ has been identified as a carrier of four diffuse interstellar bands\cite{Campbell:2015aa,Walker:2015aa}.

Despite the anticipated ubiquity of PAHs and fullerenes in space, the mechanisms for formation and growth of aromatic systems in these environments remain largely unknown. Several experimental and theoretical studies on the growth of PAHs and fullerenes have revealed a number of bottom-up processes where carbon rings are formed through the sequential addition of small hydrocarbon building blocks such as CH$_3$ and C$_2$H$_2$\cite{Richter:2000aa,Gerasimov:2009aa,Jager:2009aa,Shukla2011369,Carpentier:2013aa,Contreras:2013aa,Peverati:2016aa,Bera:2015aa,Peverati:2014aa}. However, the applicability of such chemical processes to interstellar environments is limited by low densities and fragile reaction intermediates, which are likely to photodissociate before they may grow further in the harsh environment of the ISM\cite{Mebel:2008aa}. Instead, top-down models are often used to predict the formation of fullerene cages in the ISM through the decay of larger carbon-based complexes\cite{doi:10.1021/jp061173z}, e.g.\ large PAHs (the formation paths of which also are poorly known), when these are processed by energetic photons\cite{Berne:2012aa,berne2015,Zhen:2014aa}, e.g.\ by the intense UV radiation fields in planetary nebulae, or particles\cite{Micelotta:2010aa}, e.g.\ by energetic atoms/ions in supernova shockwaves.


Recently it has been shown that keV ions colliding with loosely bound clusters of PAHs or fullerenes can induce molecular growth within these clusters\cite{PhysRevLett.110.185501,:/content/aip/journal/jcp/139/3/10.1063/1.4812790,PhysRevA.89.062708,Delaunay:2015aa,Gatchell:2016aa}. This growth is mainly driven by the prompt fragmentation of molecules in cluster when the impacting projectile ion deposits a large amount of energy and momentum to individual atoms through nuclear scattering processes\cite{Gatchell:2016aa}. When enough energy is transferred in such processes, atoms can be knocked out from the individual molecules in the cluster, creating secondary projectiles (the atoms that have been knocked out may cause further damage) and fragments with dangling bonds. These remaining molecular fragments are often highly reactive and may form covalent bonds with neighboring molecules in the cluster on sub-picosecond timescales. That is, well before the excited cluster dissociates\cite{Gatchell:2016aa}. The surrounding molecules are important here as they provide fuel for the reactions and dissipate some of the excess energy. This helps to cool the reaction products and protects them from secondary statistical fragmentation\cite{Gatchell:2016aa}. 

The elastic nuclear scattering that initiates reactions in PAH and fullerene clusters is often the main mechanism for energy transfer when atoms, ions, or molecules collide at energies below a few keV (the exact limit varies depending on the projectile)\cite{:/content/aip/journal/jcp/140/22/10.1063/1.4881603,Gatchell:2016aa}. At higher collision energies, energy is mainly transferred through inelastic electronic scattering processes that excites the electrons in the target in a similar way as in photo-absorption processes\cite{:/content/aip/journal/jcp/140/22/10.1063/1.4881603,Gatchell:2016aa,Maclot:2016aa}. In most cases, most of the energy that is transferred to the target by either mechanism will be redistributed, e.g.\ by internal conversions from electronically to vibrationally excited states, over all available internal degrees-of-freedom. Internally hot molecules will then typically cool by dissociating through their lowest energy pathways\cite{Gatchell:2016aa} and/or by emitting photons\cite{PhysRevLett.110.063003,1989ApJS...71..733A}, typically producing at most one reactive product. The prompt knockout of atoms on the other hand is a process unique to energy transfer through nuclear scattering by heavy particles and often leads to the formation of \emph{multiple} highly reactive fragments\cite{Gatchell:2016aa}. In addition to the reactions in PAH and fullerene clusters\cite{PhysRevLett.110.185501,:/content/aip/journal/jcp/139/3/10.1063/1.4812790,PhysRevA.89.062708,Delaunay:2015aa,Gatchell:2016aa}, knockout driven fragmentation has been detected in experiments with isolated C$_{60}$ anions\cite{Larsen:1999aa,Tomita:2002aa}, C$_{60}$ cations\cite{Gatchell2014260}, and PAH cations\cite{PhysRevA.89.032701,Gatchell2014260}. The balance between nuclear and electronic scattering can be studied choosing different projectiles ions and energies---slow and heavy projectile ions interact mainly through nuclear scattering, and fast and light ions interact mainly through electronic scattering. Furthermore, the importance of both processes and the reaction dynamics may also depend on the structure and size of the irradiated molecule, and whether the molecule is isolated or in a cluster. It has for example been shown that radical cations such as C$_8$H$_8^{\bullet+}$  can be formed through electron impact ionization of clusters of small hydrocarbon chain molecules such as acetylene, C$_2$H$_2$\cite{Momoh:2011aa}. Thus, it appears that simple ionization may also induce chemical reactions and isomerization processes \cite{Kocisek:2013aa,Nagy:2011aa,Franceschi:2002aa,Leidlmair:2012aa}.

Here we investigate if impulse driven (knockout) reactions like those observed in PAH and fullerene clusters can drive the growth of small hydrocarbon chains into larger structures and to form aromatic molecules from non-aromatic ones. In the next two sections we present the experimental setup and theoretical methods that we use here. We then move on to show and compare experimental and theoretical results for 3\,keV Ar$^+$ ions colliding with isolated 1,3-butadiene (C$_4$H$_6$) molecules and their loosely bound clusters. We find that these collisions result in a wealth of molecular growth products that could include aromatic rings.

\section{Experimental Details}

\begin{figure}[t]
  \centering
    \includegraphics[width=0.75\columnwidth]{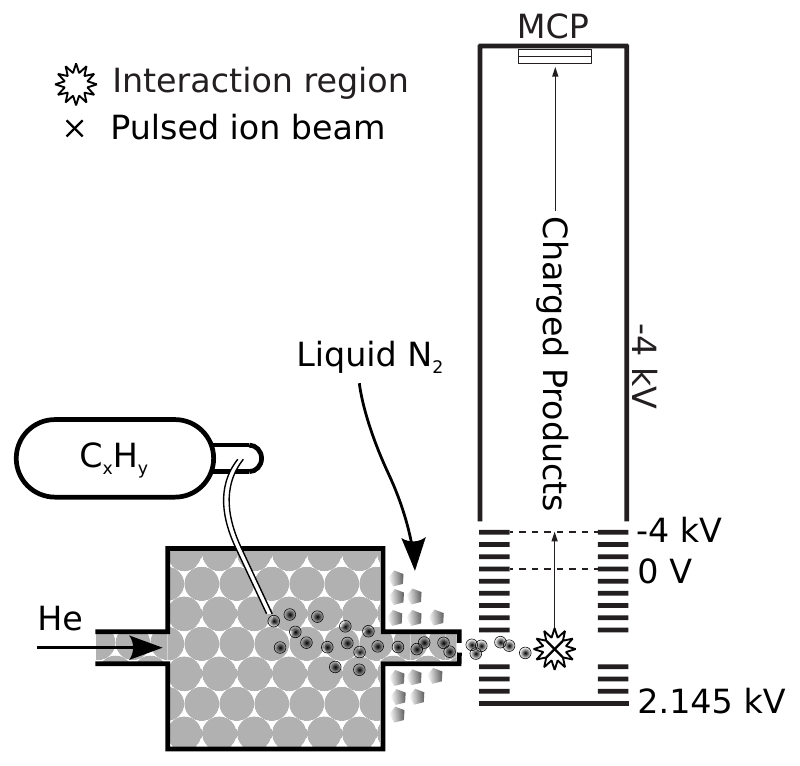}
  \caption{Schematic of the experimental setup at the GANIL facility in Caen. A neutral beam of either clusters or molecules enters the interaction region from the left and the 3\,keV Ar$^+$ ion beam (cross symbol) enters perpendicular to this (into the page). Positively charged reaction products and fragments are analyzed with the time-of-flight mass spectrometer (to the right and top).}
  \label{fig:aribe}
\end{figure}

The experimental part of this work was performed at the ARIBE low energy ion facility at GANIL in Caen, France, using the crossed beam apparatus\cite{Bergen:1999aa} shown schematically in Figure \ref{fig:aribe}. The butadiene target was supplied from a bottle which fed gas into a cluster aggregation source. In experiments involving molecular clusters, the cluster source is cooled with liquid nitrogen at 77\,K and a He buffer gas is injected to aid the cooling of the butadiene gas, which then condenses into clusters with a broad log-normal size distribution\cite{Bergen:1999aa}. It is not possible to isolate individual cluster sizes with the present setup. Beams of isolated molecular (monomer) targets are formed in the same way, but with the cluster source operating at ambient temperature. The neutral clusters or molecules, depending on which of the targets that is prepared, leave the source through a nozzle and pass a differential pumping stage before entering the interaction region of the experiment where they cross with the keV ion beam. The 3\,keV Ar$^+$ ions are produced by means of an ion gun and standard equipment to yield 3.0\,{\textmu}s long pulses.

The interaction region is placed in the extraction zone in a Wiley-McLaren time-of-flight mass spectrometer\cite{Chandezon:1994aa}. After the ion-beam pulse has left this region, the extraction voltage of the mass spectrometer is applied and positively charged products are accelerated into the field free region. The charged products then hit a position sensitive 80\,mm microchannel plate (MCP) detector (biased to $-10$\,kV) at the end of the 1 meter drift tube and the time-of-flight of each charged product (with respect to the time when the extraction voltage is switched on) is recorded with ns time resolution. 

\section{Computational Details}
Following the approach successfully used to model collision induced molecular growth in PAH clusters \cite{Delaunay:2015aa}, we have performed classical Molecular Dynamics (MD) simulations of neutral Ar atoms colliding with 1,3-butadiene clusters. We used the AIREBO (Adaptive Intermolecular Reactive Empirical Bond Order) potential\cite{Stuart:2000aa} to describe the chemical interactions between C and H atoms in the molecules as well as non-bonding dispersive forces between molecules. The AIREBO potential is a reactive potential that allows for bonds to be formed and broken dynamically during the simulation based on the bond order and type of each atom and its neighbors. The interactions between the projectile and the clusters were described using the ZBL (Ziegler-Biersack-Littmark) potential \cite{zbl_pot_book}, a screened Coulomb potential for modeling nuclear scattering processes. These classical force fields do not contain an explicit description of charge or polarization. However, previous results have shown that these methods still produce results that are in good agreement with experiments and higher level theory where charged reaction products are studied\cite{Delaunay:2015aa,PhysRevA.89.062708,PhysRevLett.110.185501}. We used the LAMMPS package\cite{Plimpton:1995aa} and the forcefields as defined therein for the present simulations.


\begin{figure}[t]
  \centering
    \includegraphics[width=0.75\columnwidth]{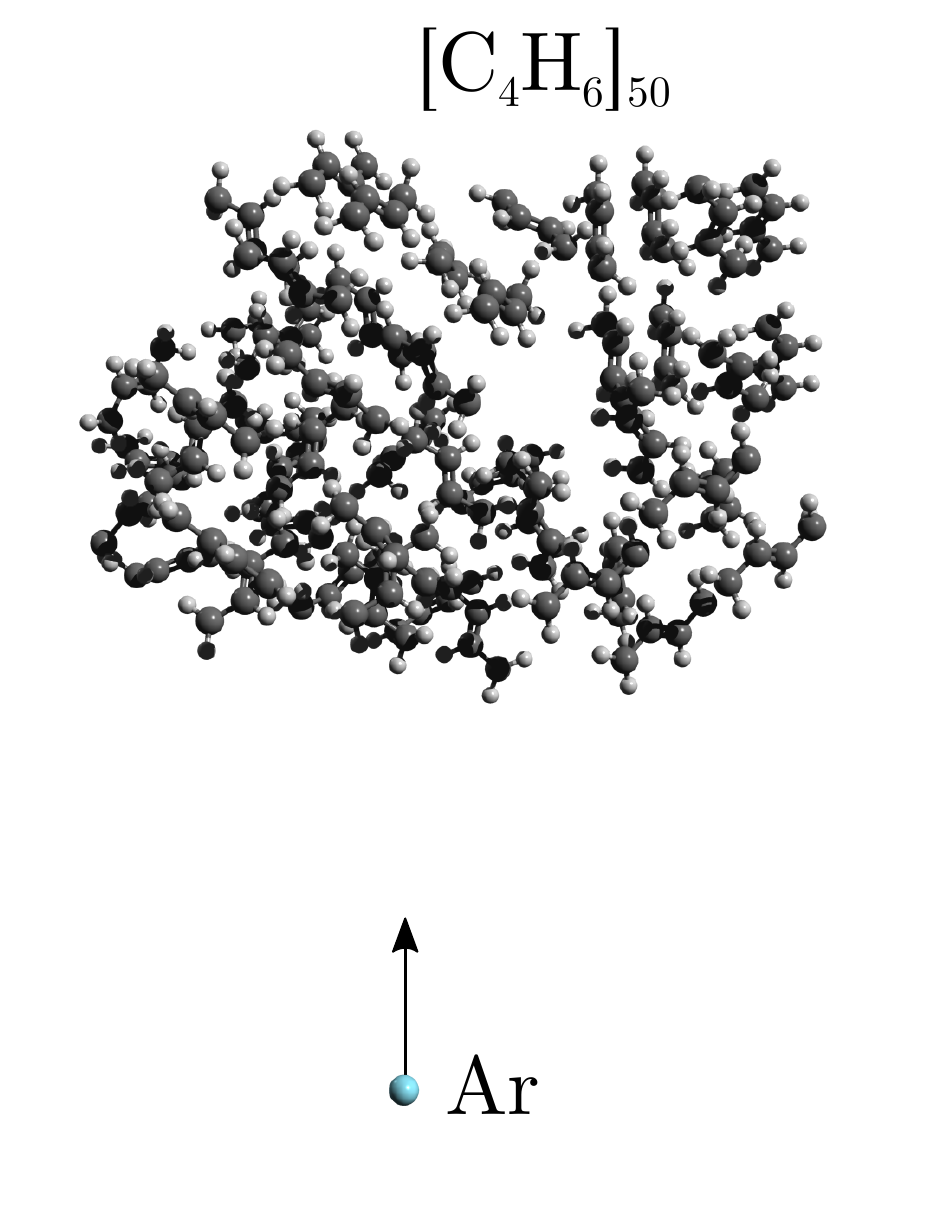}
  \caption{An example of an Ar atom which collides with a randomly oriented [C$_4$H$_6$]$_{50}$ cluster with a randomly selected impact parameter. The cluster geometry has been optimized using the AIREBO potential (see text).}
  \label{fig:sims}
\end{figure}

The simulations were performed by first generating a system of 50 closely separated C$_4$H$_6$ molecules with random orientations. We then optimized the geometry of the cluster using the AIREBO potential. The resulting cluster has a compact structure and a binding energy (from dispersion forces) of 190\,meV per molecule. This cluster geometry (shown in Fig.\ \ref{fig:sims}) was used for all of the classical MD simulations performed here. For each simulation run, the cluster was positioned with its center-of-mass at the origin of a three-dimensional coordinate system and was then randomly rotated around this origin. The Ar projectile atom was fired with a randomly selected impact parameter (with the respect to the origin of the coordinate system) towards the cluster from a position 20\,{\AA} away. The collision occurs within the first few fs of the simulation and the system was followed for a total simulation time of 10\,ps (using a time step of $5 \times 10^{-17}$\,s). At the end of a simulation the positions of all the atoms were analyzed to determine which covalent bonds have been broken and which new ones that have been formed. We repeated this simulation process 10,000 times to produce a mass spectrum. Figure \ref{fig:sims} shows an Ar + [C$_4$H$_6$]$_{50}$ system at the first time step of a simulation run. We have tested the simulation with different cluster sizes. The size has little effect on the types of reactions that take place in the clusters---i.e.\ on the simulated fragment mass spectrum---but as expected larger clusters give greater reaction probabilities for geometrical reasons.
 
In addition, we have performed Density Functional Theory (DFT) calculations to determine the relative stability of the different isomers of a given molecular species (molecular growth product) that is formed in the collisions. The complete sets of the very large number of possible isomers for each studied reaction product were generated using the Open Molecule Generator\cite{Peironcely2012}. In a first step we optimized all of these structures at the B3LYP/3-21G level of theory. The lowest energy isomers (those within 2\,eV of the lowest energy structure) were then re-optimized at the B3LYP/CC-pVDZ level of theory. For these systems we also performed a frequency analysis to ensure that real energy minima had been found. These DFT calculations were performed using the Gaussian 09 software\cite{Frisch:2009aa_long}. 

\section{Results and discussion}

\subsection{Collisions with isolated butadiene molecules}
\begin{figure}[t]
\centering
    \includegraphics[width=1\columnwidth]{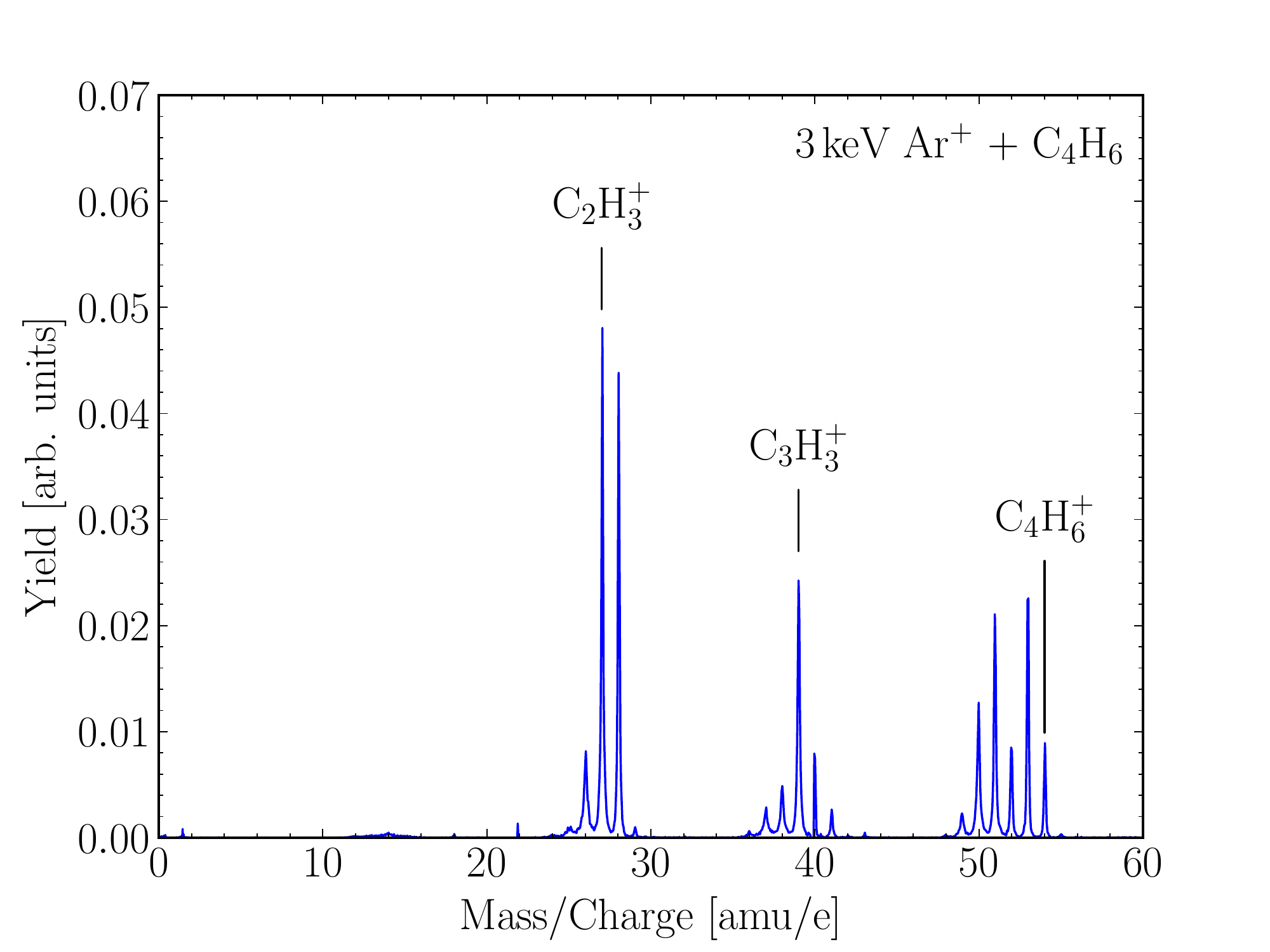}
  \caption{Ionization and fragmentation mass spectrum due to 3\,keV Ar$^+$ ions colliding with isolated butadiene (C$_4$H$_6$) molecules.}
  \label{fig:MSmono}
\end{figure}

In Figure \ref{fig:MSmono} we show a mass spectrum for 3\,keV Ar$^+$ ions colliding with \emph{isolated} butadiene molecules (by operating the cluster source at room temperature and without the buffer gas).

The fragment mass spectrum exhibits strong peaks from H-loss (without C-loss) resulting in fragments down to C$_4$H$^+$ at 49\,amu/e. We also see an odd-even effect in the number of H atoms for the C$_4$H$_x^+$ fragments with the odd-numbered fragments (which have closed electronic shells) being significantly more abundant than those with even numbers of H atoms. Other notable features are the strong peaks due to C$_2$H$_3^+$, C$_2$H$_4^+$, and C$_3$H$_3^+$ fragments. The latter is the main dissociation product observed in photodissociation \cite{Robinson:2002aa} and electron impact \cite{Linstrom:2005aa} experiments with 1,3-butadiene. However, here we observe a wider range of fragment sizes than in the photodissociation experiments\cite{Robinson:2002aa}, and more small fragments, e.g.\ C$_2$H$_3^+$, C$_2$H$_4^+$, than in the electron impact experiments\cite{Linstrom:2005aa}. These differences could be caused by the molecules having higher excitation energies in our experiments, but could also be influenced by knockout-driven fragmentation.

\subsection{Collisions with butadiene clusters}

\begin{figure}[t]
\centering
    \includegraphics[width=1\columnwidth]{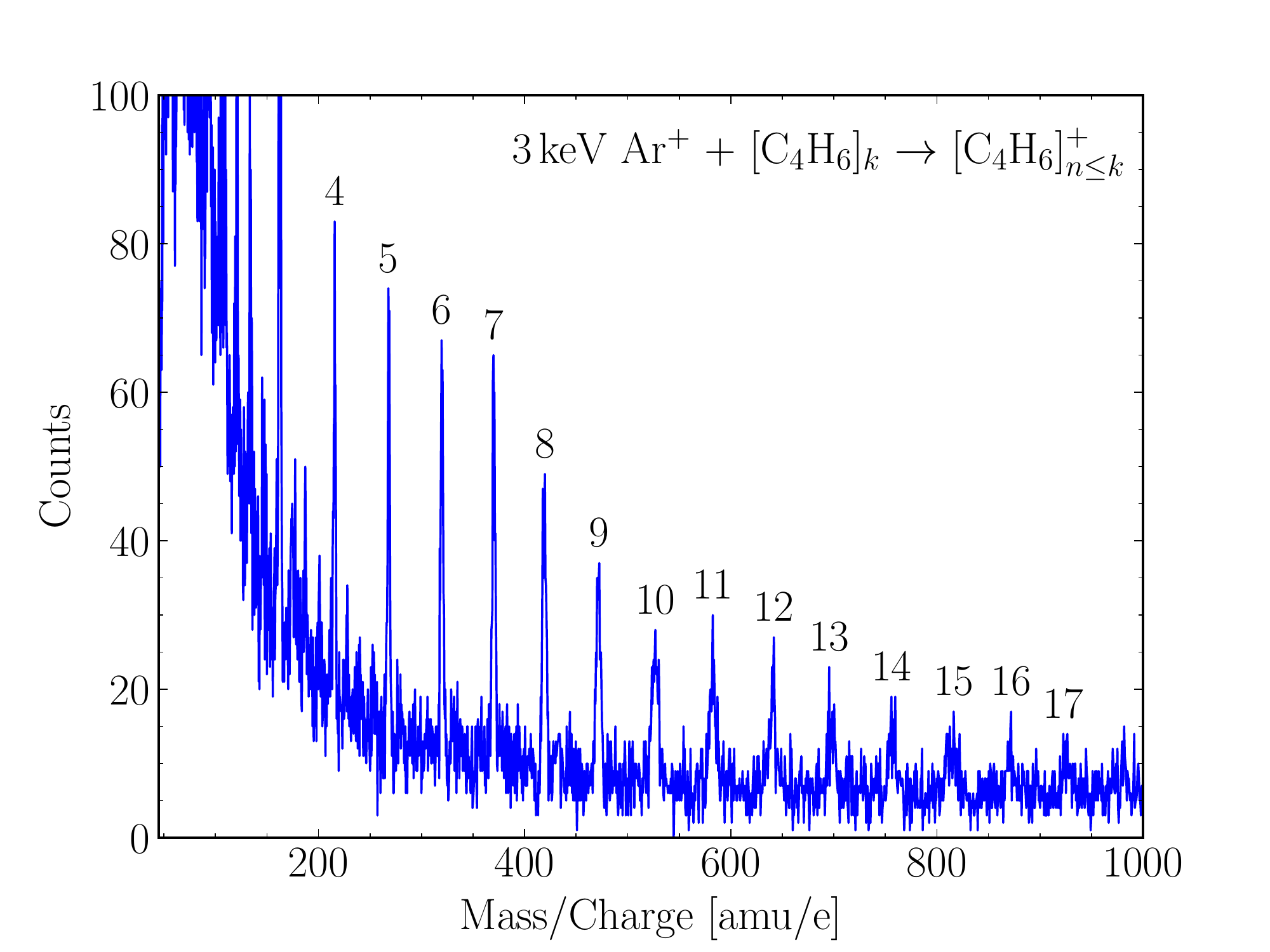}
  \caption{Overview mass spectrum of intact, singly charged butadiene clusters produced in collisions with 3\,keV Ar$^+$ projectiles. The numbers above the peaks indicate the number, $n$, of intact molecules in a given cluster size.}
  \label{fig:MSclusters}
\end{figure}

When cooling the cluster aggregation source to 77\,K the 1,3-butadiene molecules condense into weakly bound clusters. Figure \ref{fig:MSclusters} shows an overview mass spectrum of intact, singly ionized clusters ranging in sizes of up to at least 17 molecules from collisions between these neutral clusters and the 3\,keV Ar$^+$ projectiles. These charged clusters are mainly formed when the projectile ions capture an electron from the neutral clusters at large distances, leaving the charged clusters with low internal energies. Due to the low binding energy of the butadiene clusters (190\,meV in our MD simulations of large clusters, 120\,meV per molecule in DFT calculations of dimers\cite{Premkumar:2014aa}), even the low excitation energies involved when the clusters are ionized, together with other processes such as ion induced dipole interactions within in the charged clusters, could be enough to induce some cluster dissociation. This dissociation primarily takes place through the evaporation of intact molecular monomers from the charged clusters.  


\begin{figure*}[h]
\centering
    \includegraphics[width=1\textwidth]{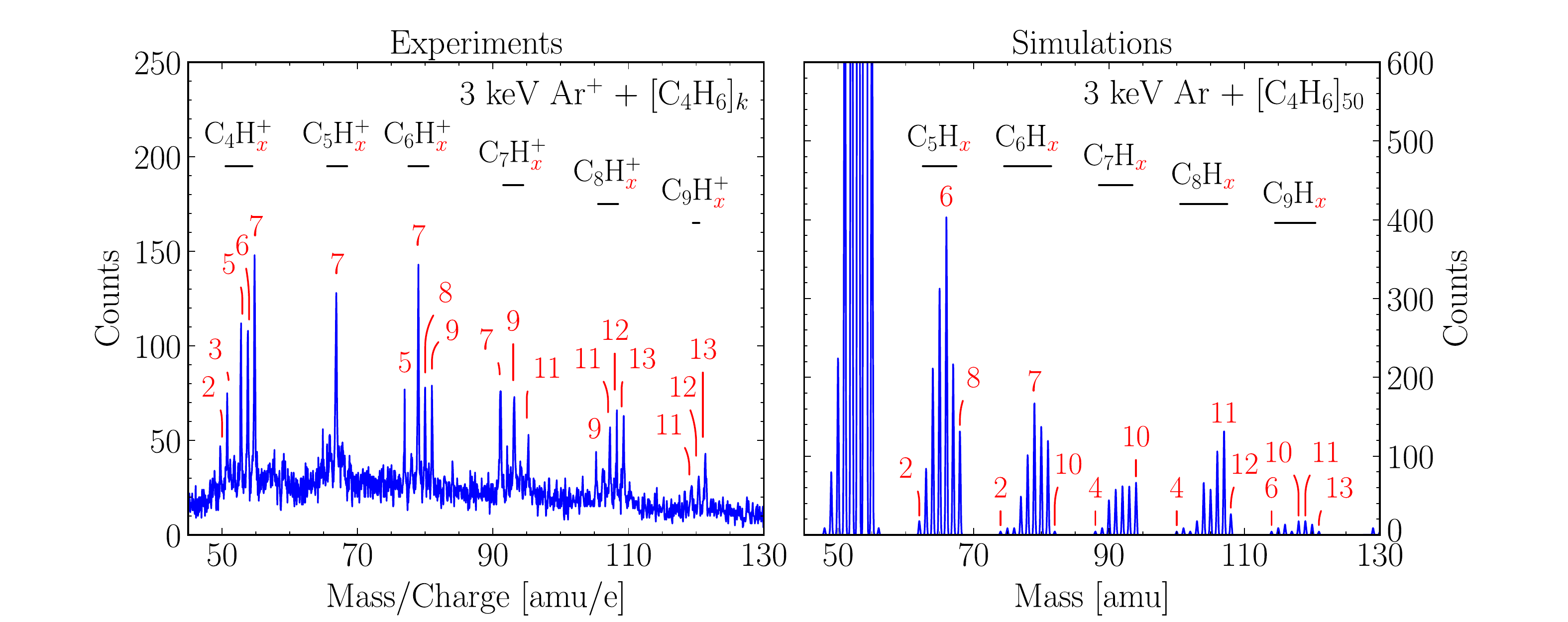}
  \caption{Total mass spectrum showing the mass range from below the intact C$_{4}$H$_{6}^+$ to products consisting of up to 9 C atoms. The left panel shows the experimental mass spectrum from 3\,keV Ar$^+$ projectiles. The right panel shows the corresponding mass spectrum from the present classical MD simulations. The peaks are grouped by the number of C atoms and labelled (in red) by the number of H atoms they contain. In the simulated mass spectrum, the lowest mass, highest mass, and most abundant growth products for each number of C atoms are labelled. Note that the simulations do not distinguish between neutral and charged reaction products. This is the main reason for the large intensities of C$_4$H$_x$ products in the simulations---as many collisions yield large numbers of neutral C$_4$H$_x$ due to cluster fragmentation.}
  \label{fig:MSgrowth}
\end{figure*}

In the left panel of Figure \ref{fig:MSgrowth} we present a zoom-in of the same mass spectrum as in Fig.\ \ref{fig:MSclusters}, focusing on the region from the intact molecular cation up to a mass above that of the molecular dimer. We see that a wide range of products with mass-to-charge ratios greater than C$_4$H$_6^+$ (54\,amu/e) are detected. The most prominent peak in this spectrum corresponds to the protonated butadiene cation, C$_4$H$_7^+$. In addition, we measure peaks over a wide range of masses from molecules consisting of between 5 and 9 C atoms, each with different distributions of the number of H atoms. The C$_5$H$_x^+$ group in particular stands out with only C$_5$H$_7^+$ giving a strong signal. For C$_6$H$_x^+$ and C$_7$H$_x^+$ products there are broader distributions in the number of H atoms, although molecules with odd numbers of H atoms mostly are significantly more abundant than those with even numbers. The only exception to this is the C$_6$H$_8^+$ peak, which is comparable to the C$_6$H$_9^+$ peak in intensity. In the group of products with 8 C atoms, the peak from C$_8$H$_{12}^+$ stands out among the even-numbered molecules although this product could also be due to intact, singly charged van der Waals dimers [C$_4$H$_6$]$_2^+$. For the products with 9 C atoms, the odd-even effect for the number of H atoms is not observed.

\begin{table*}[t]

\small
  \caption{Lowest energy cyclic and open isomers for C$_5$H$_x^+$, where $x=6,7,8$, and C$_6$H$_x^+$, where $x=5,6,7,8,9$, with zero-point corrected energies relative to the global minima for each size at the B3LYP/CC-pVDZ level of theory (see text).}
  \label{tbl:dft}
  \begin{tabular*}{1\textwidth}{@{\extracolsep{\fill}}lccc|ccccc}
 {}    & C$_5$H$_6^+$ & C$_5$H$_7^+$ & C$_5$H$_8^+$ & C$_6$H$_5^+$ &C$_6$H$_6^+$ & C$_6$H$_7^+$ & C$_6$H$_8^+$ & C$_6$H$_9^+$ \\
    \hline
   \multirow{5}{*}{Cyclic} & 0\,eV & 0\,eV & 0.16\,eV&  0\,eV & 0\,eV & 0\,eV& 0\,eV & 0\,eV \\
       & \includegraphics[height=1.5cm]{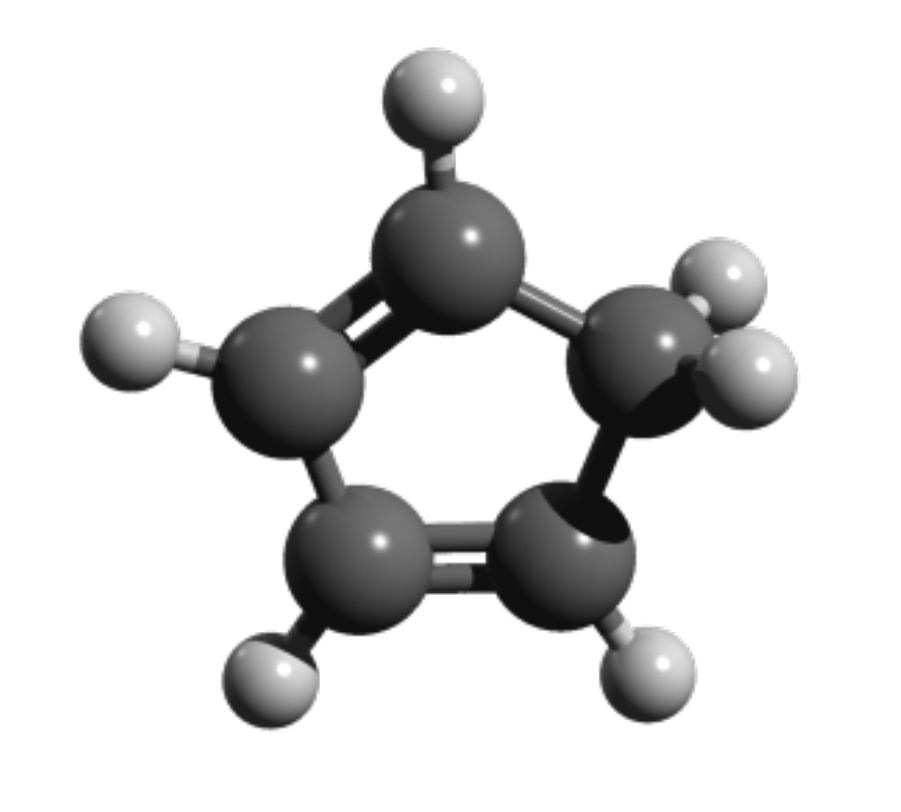}  & \includegraphics[height=1.5cm]{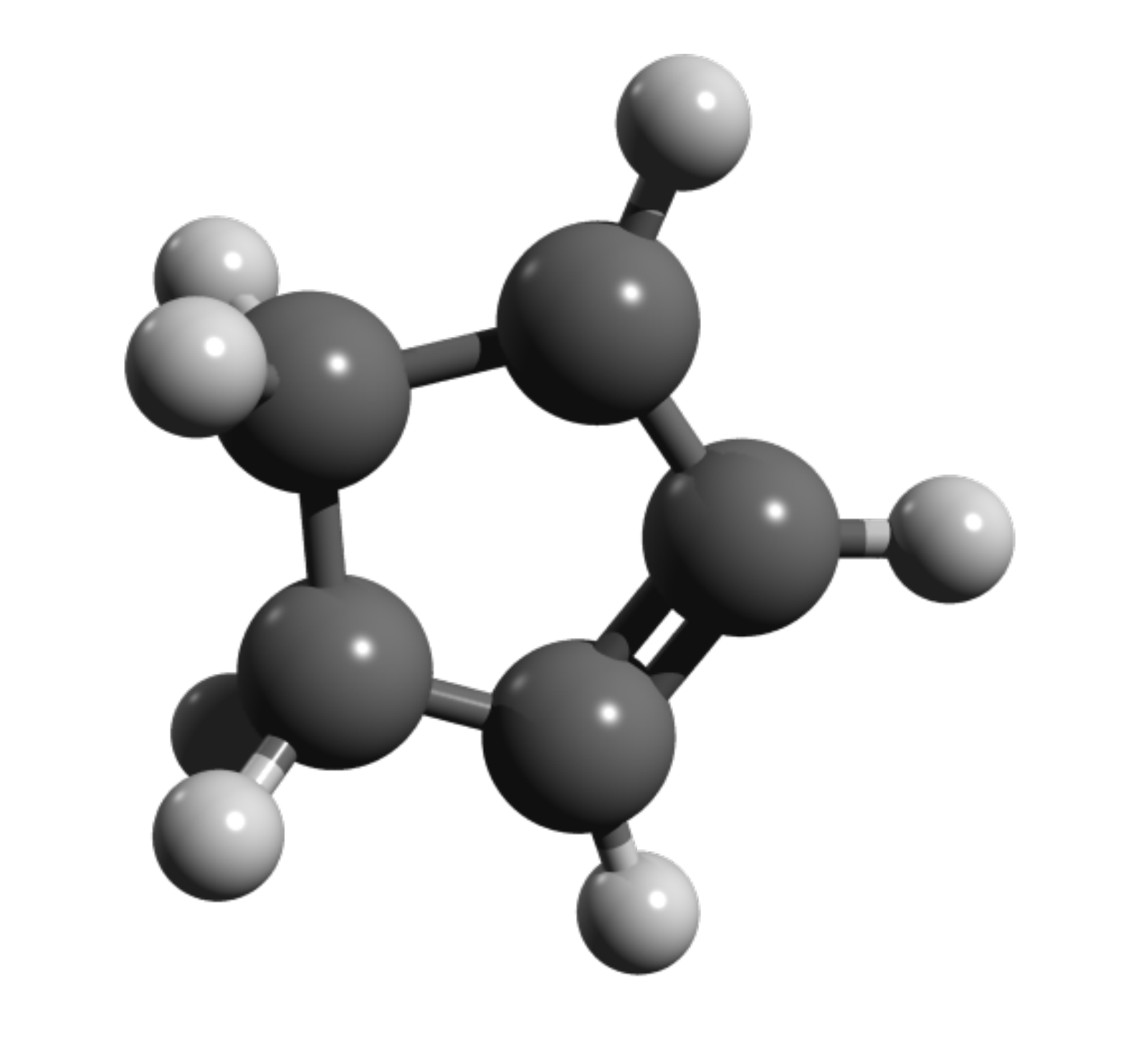}  &  \includegraphics[height=1.5cm]{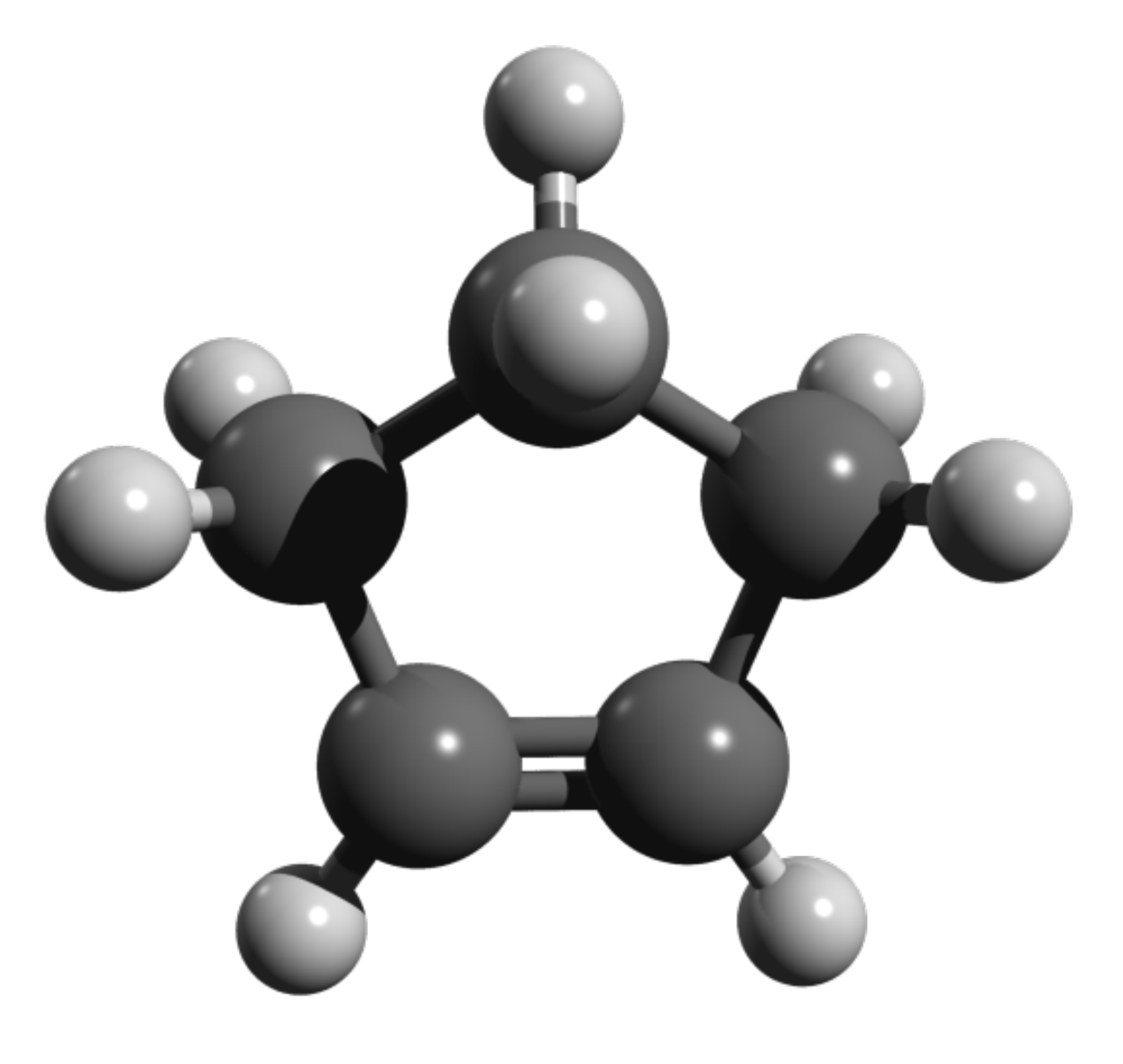} & \includegraphics[height=1.5cm]{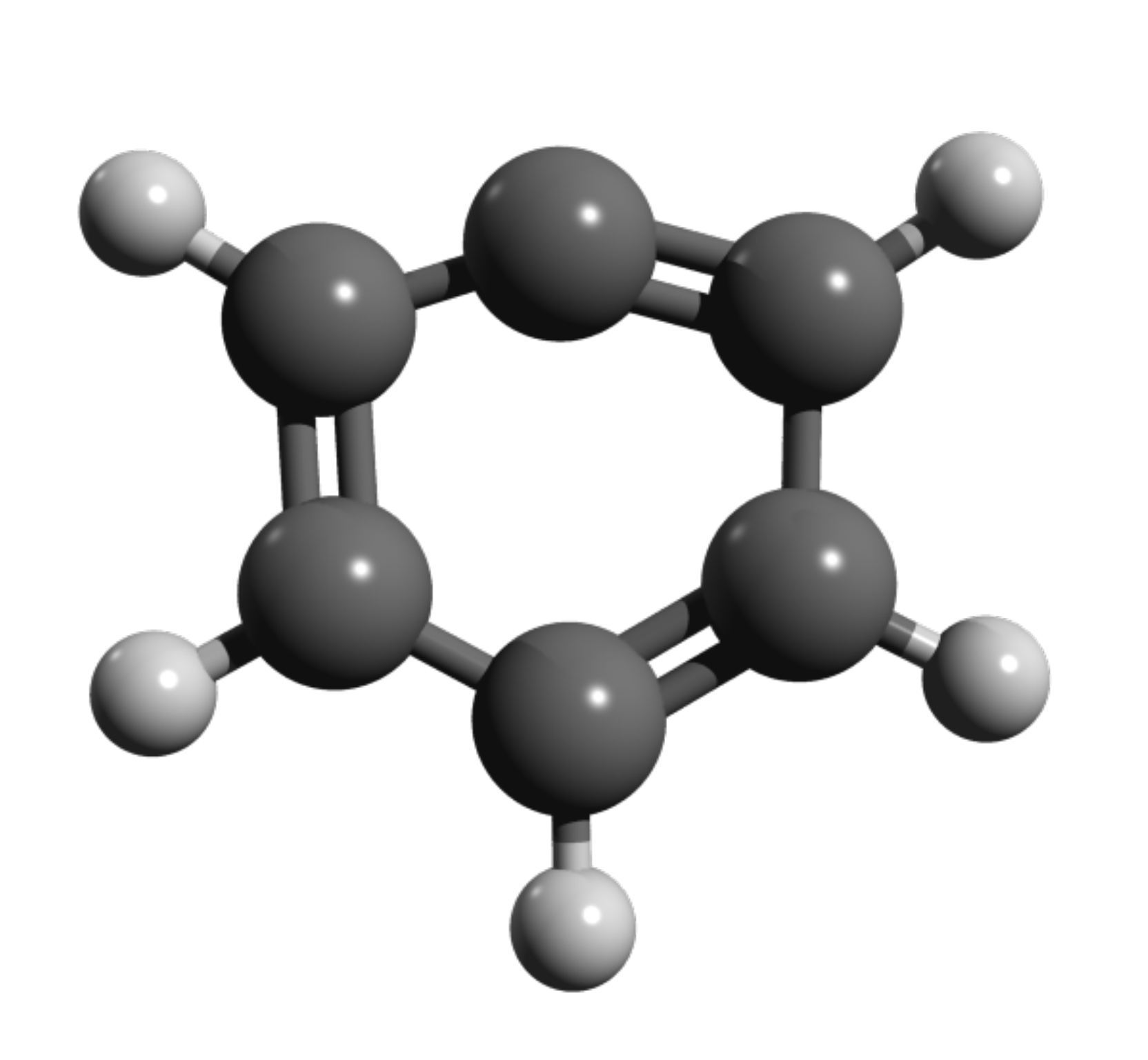}  & \includegraphics[height=1.5cm]{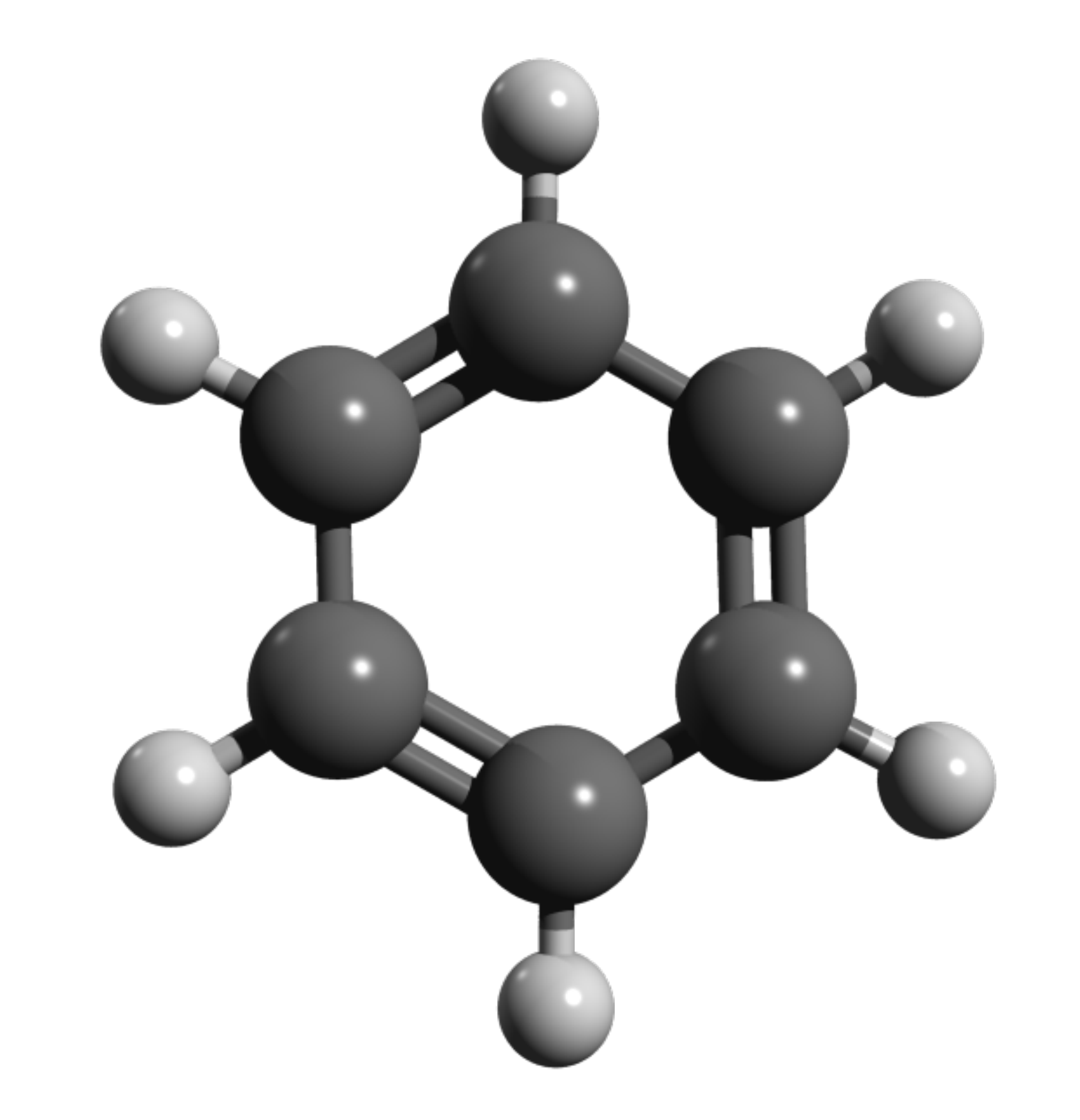}  &  \includegraphics[height=1.5cm]{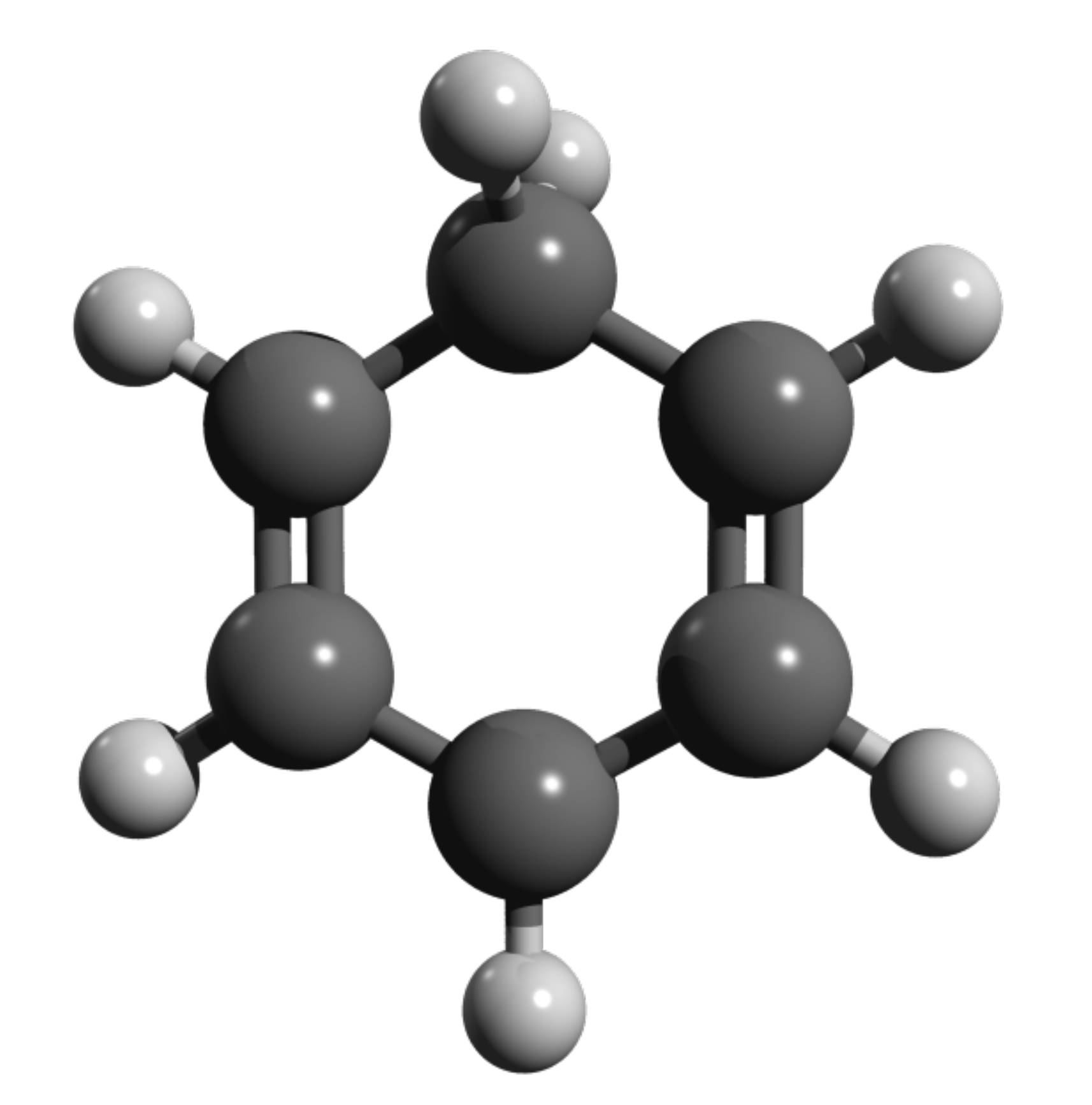}& \includegraphics[height=1.5cm]{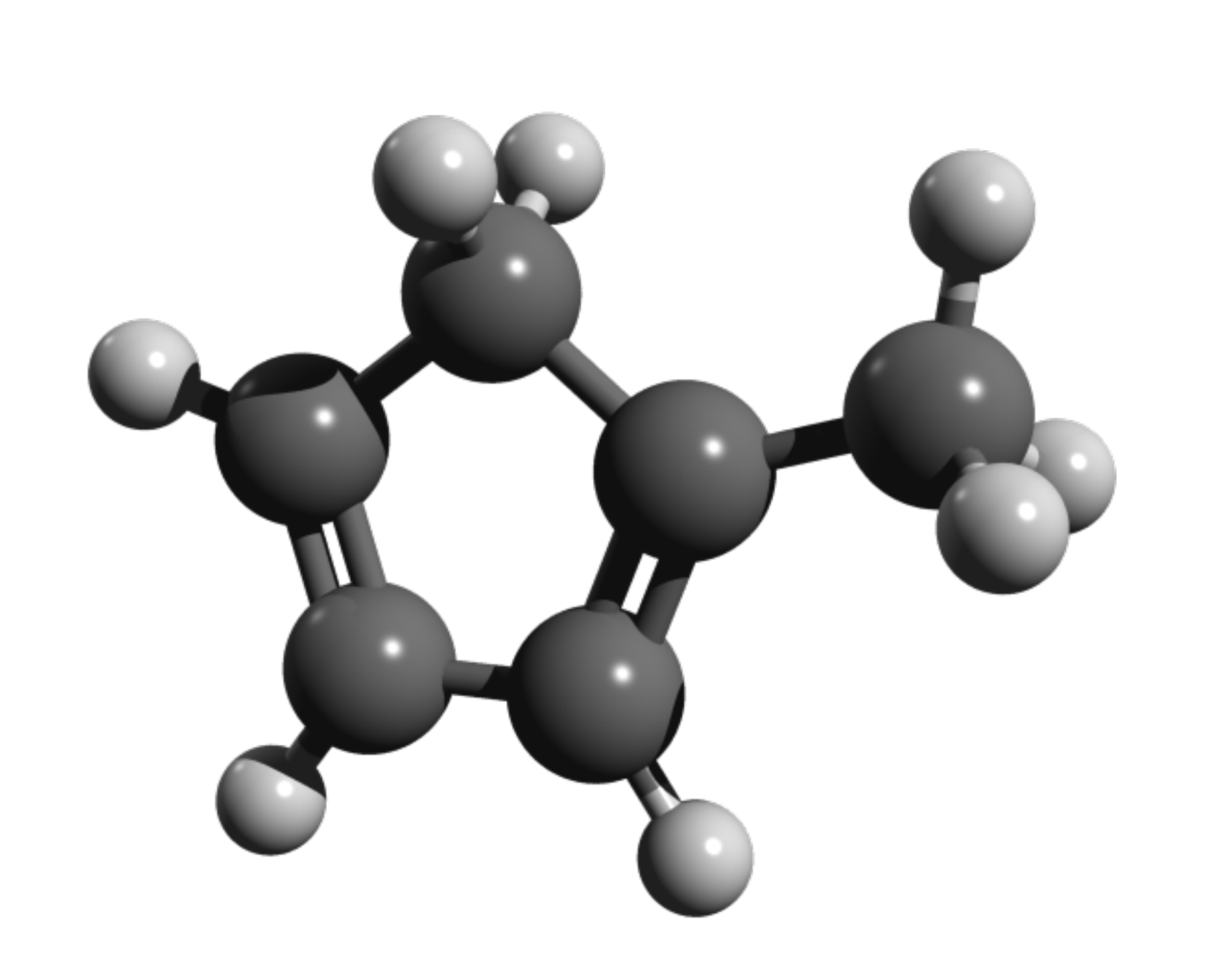}  &  \includegraphics[height=1.5cm]{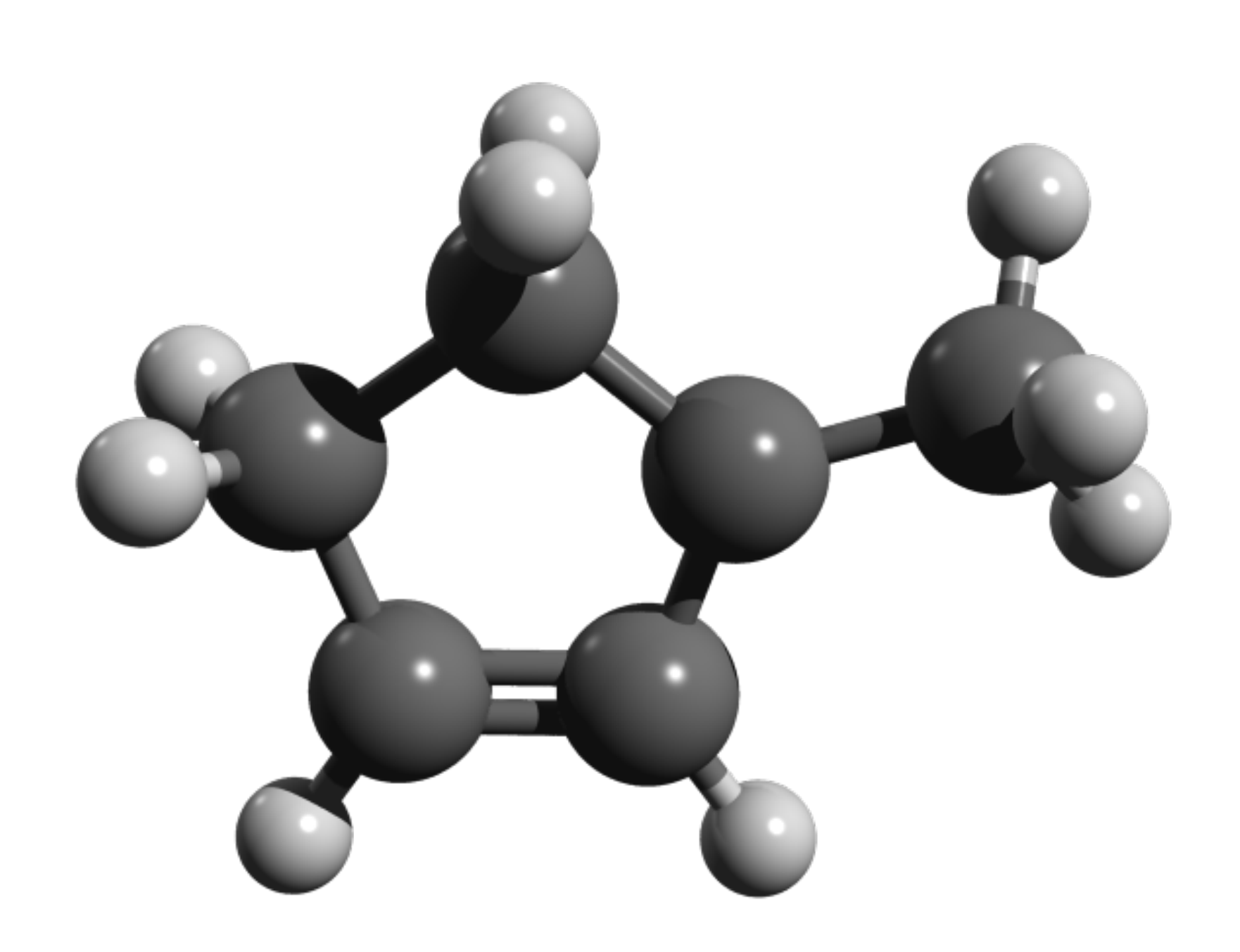} \\
\hline
     \multirow{4}{*}{Open} & 1.19\,eV & 0.66\,eV & 0\,eV & 0.84\,eV & 1.79\,eV & 1.55\,eV& 0.56\,eV & 0.72\,eV \\
        & \includegraphics[height=1cm]{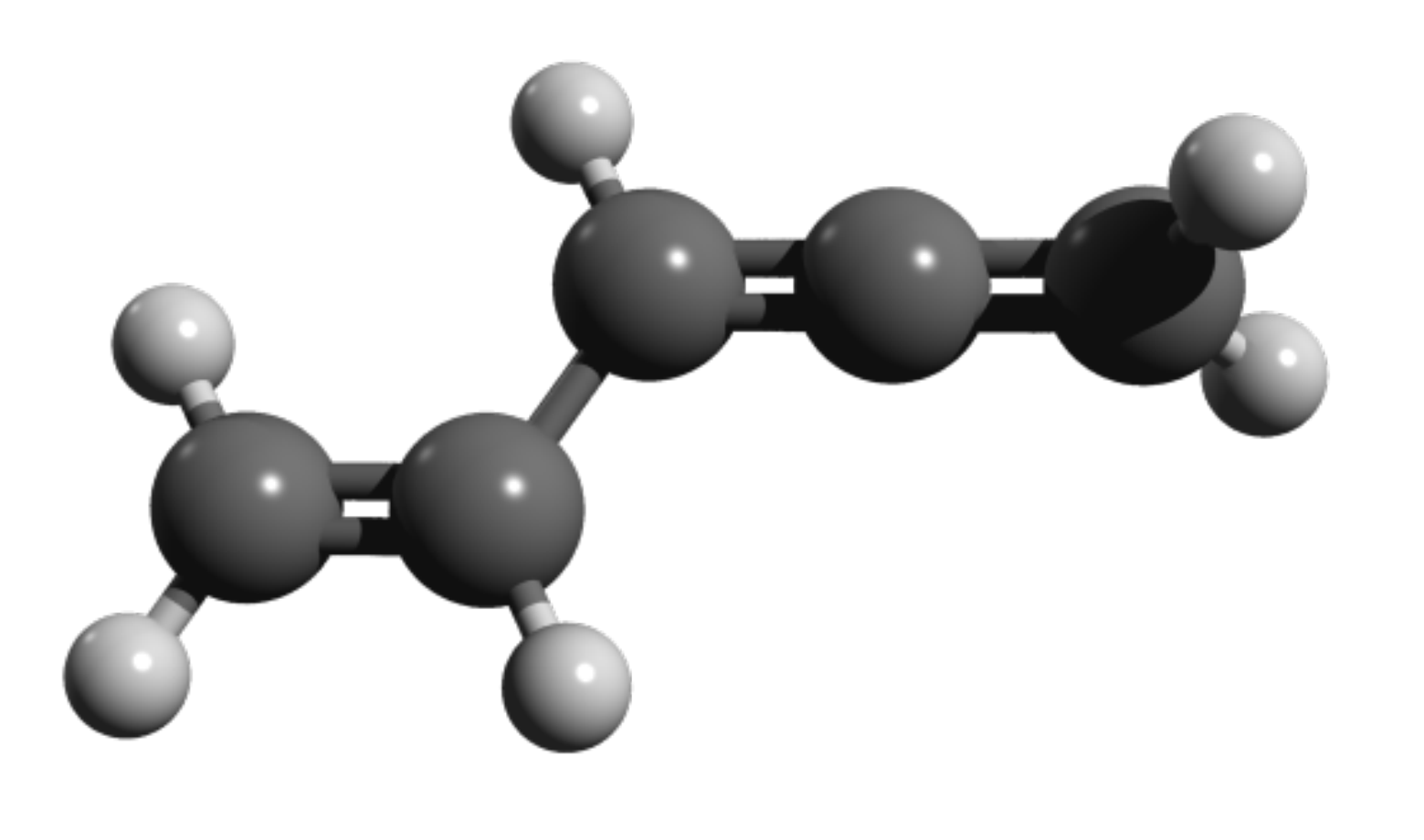}  & \includegraphics[height=1cm]{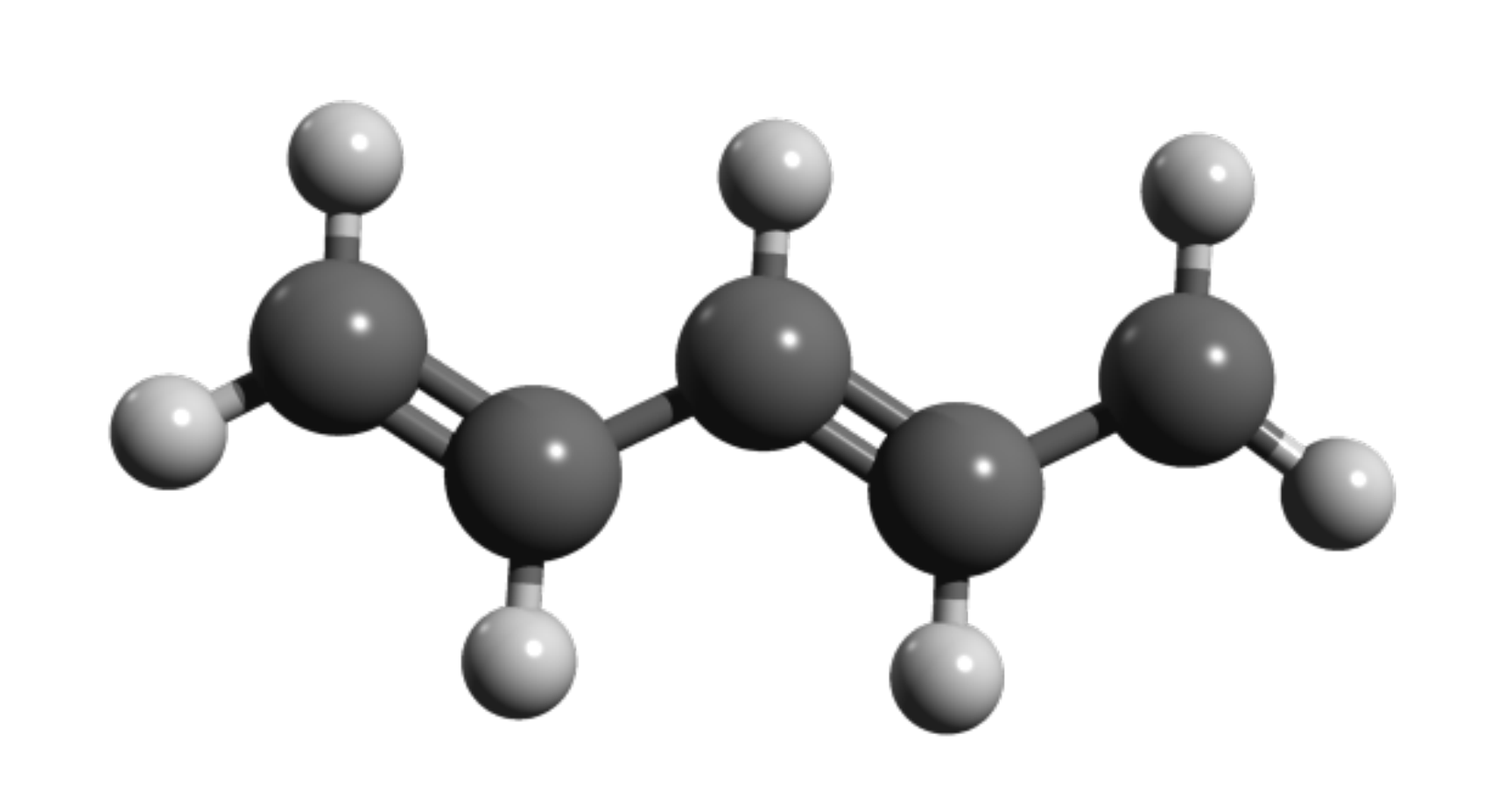} & \includegraphics[height=1cm]{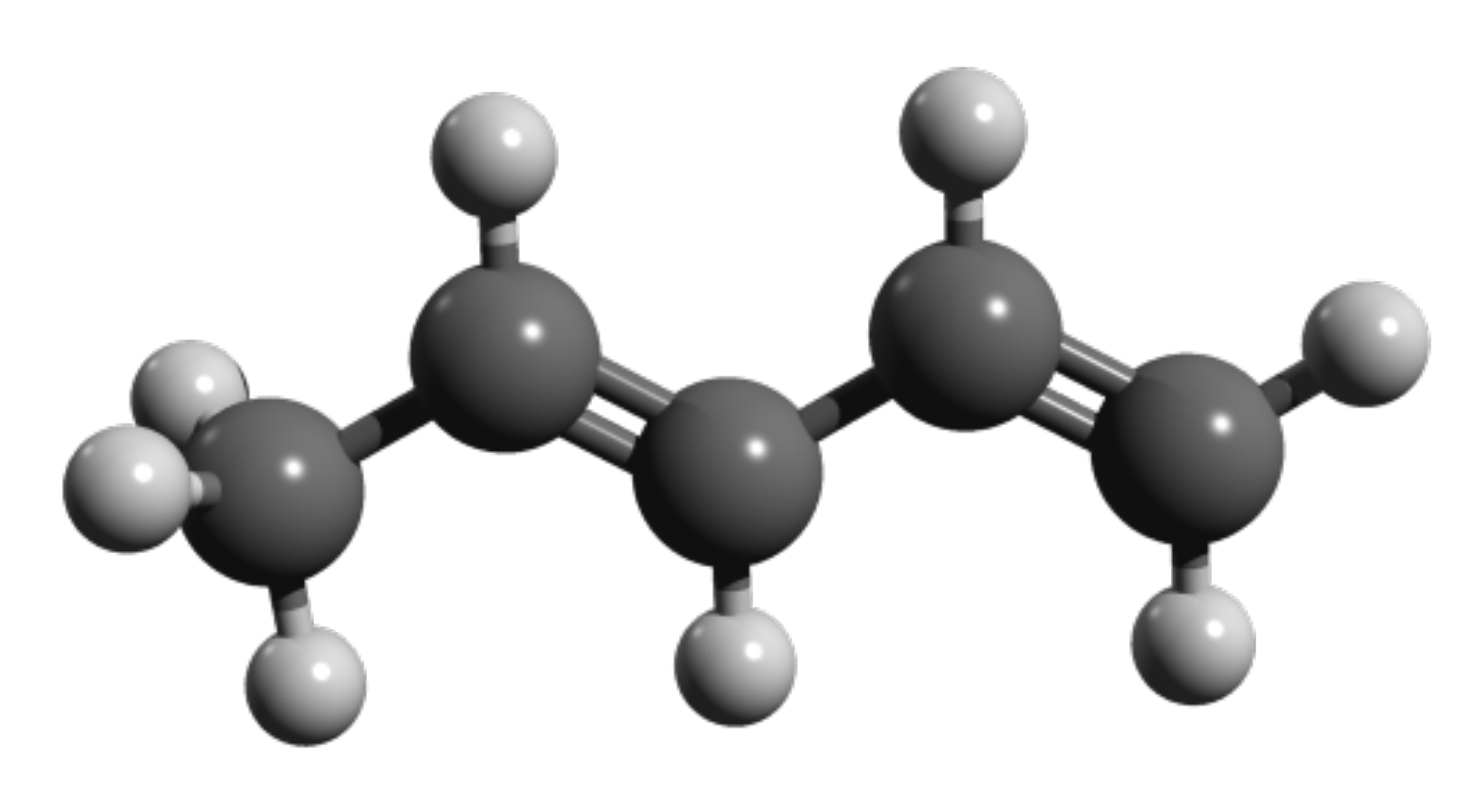} &   \includegraphics[height=1cm]{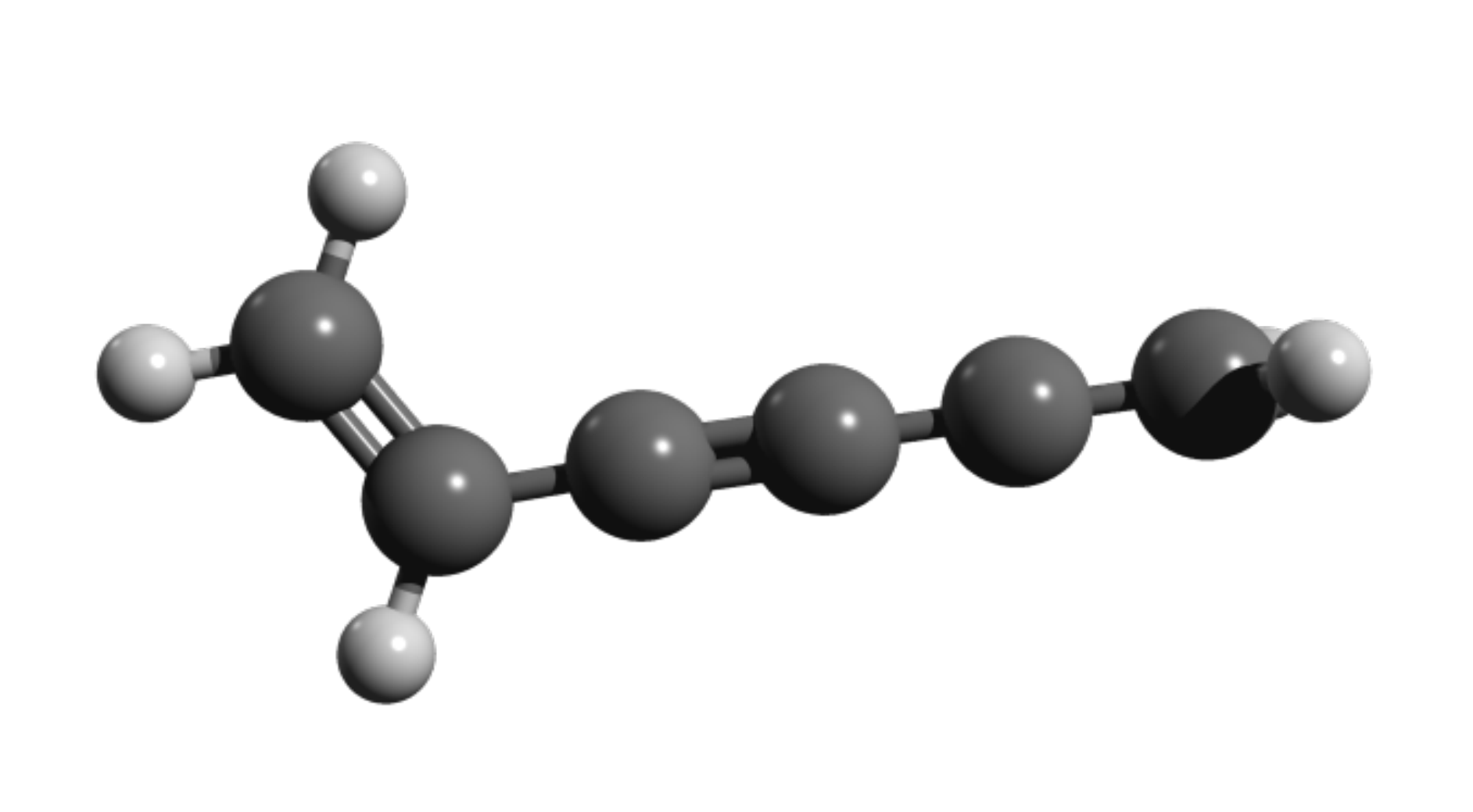}  & \includegraphics[height=1cm]{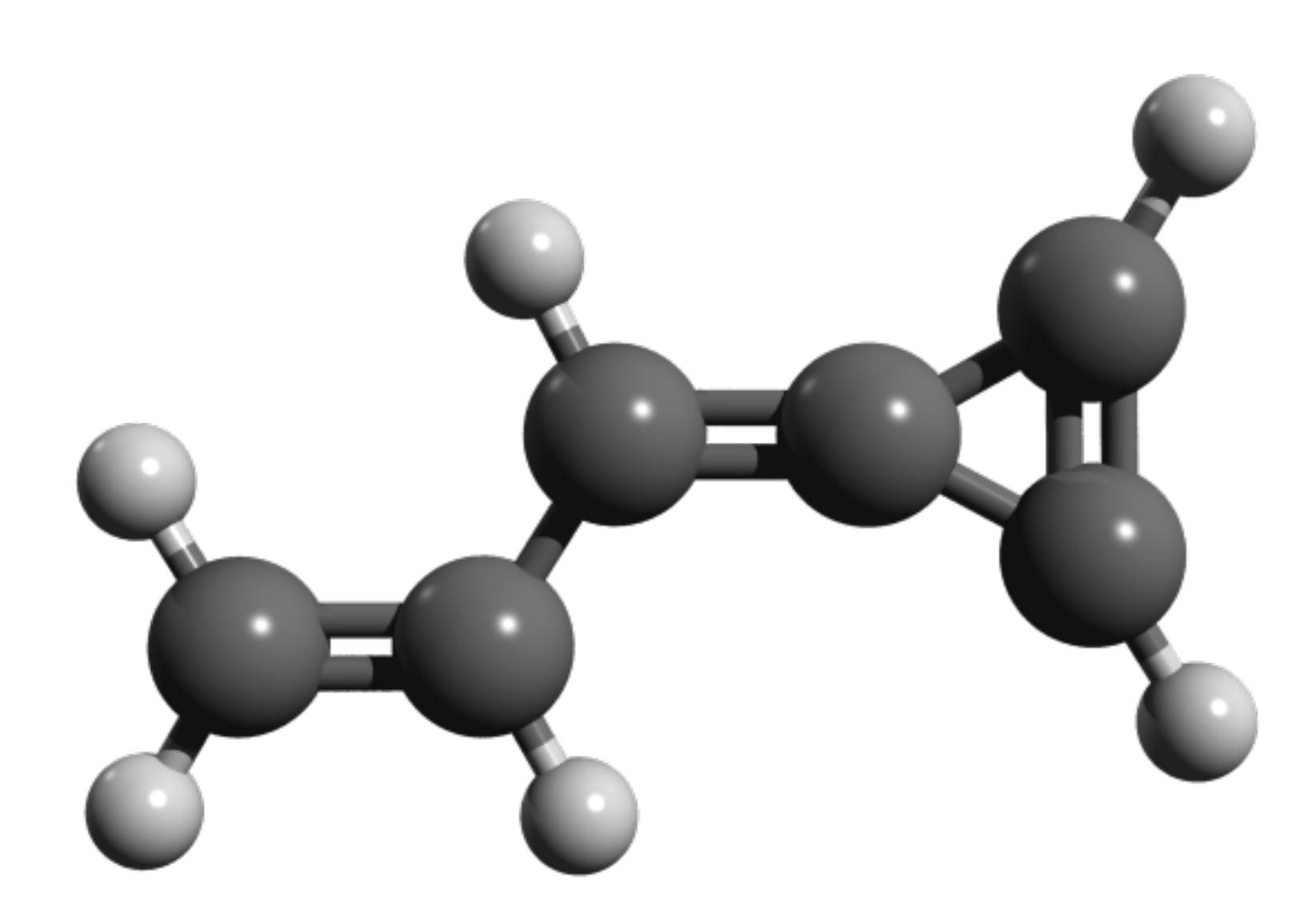} & \includegraphics[height=1cm]{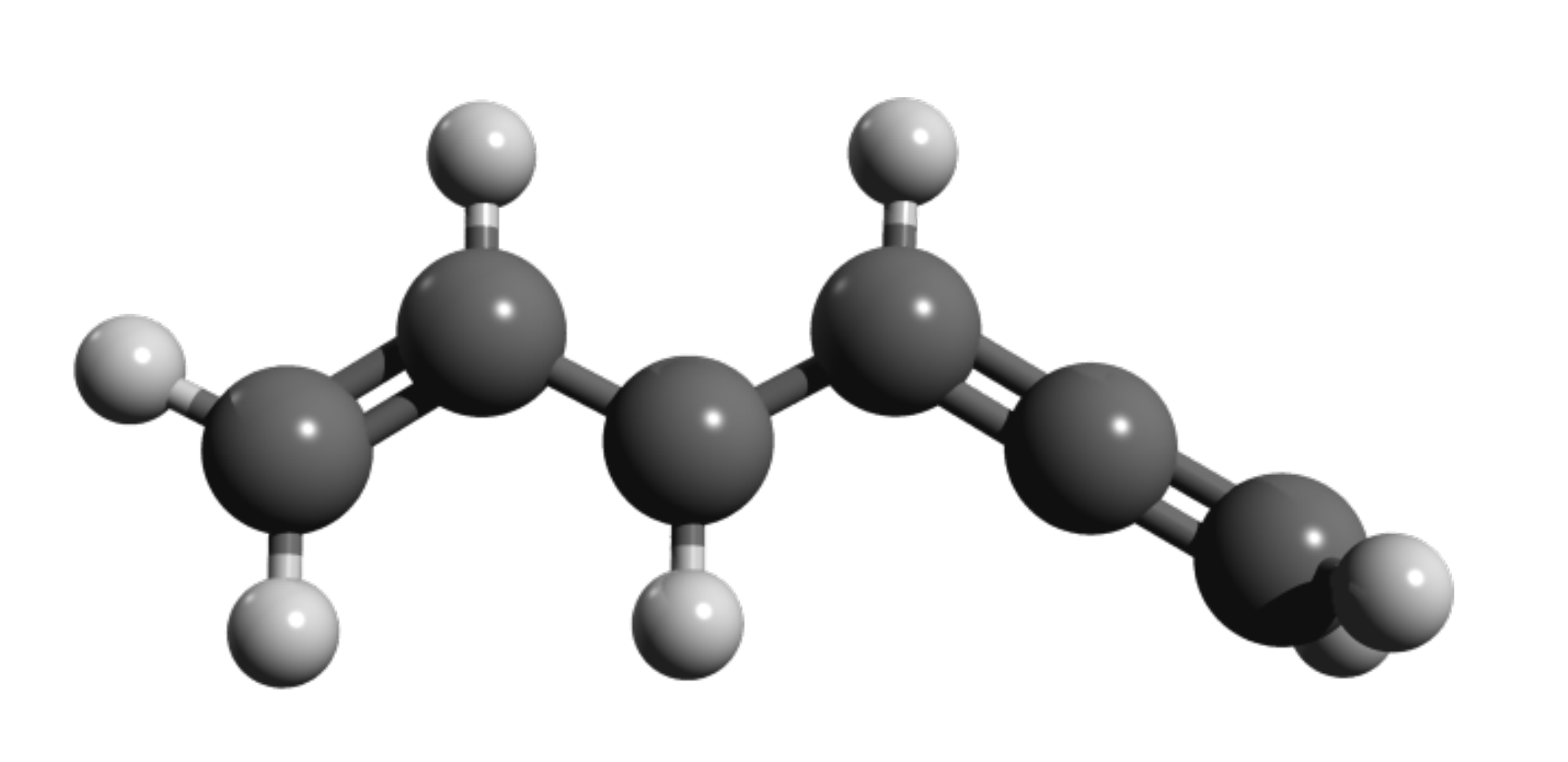}& \includegraphics[height=1cm]{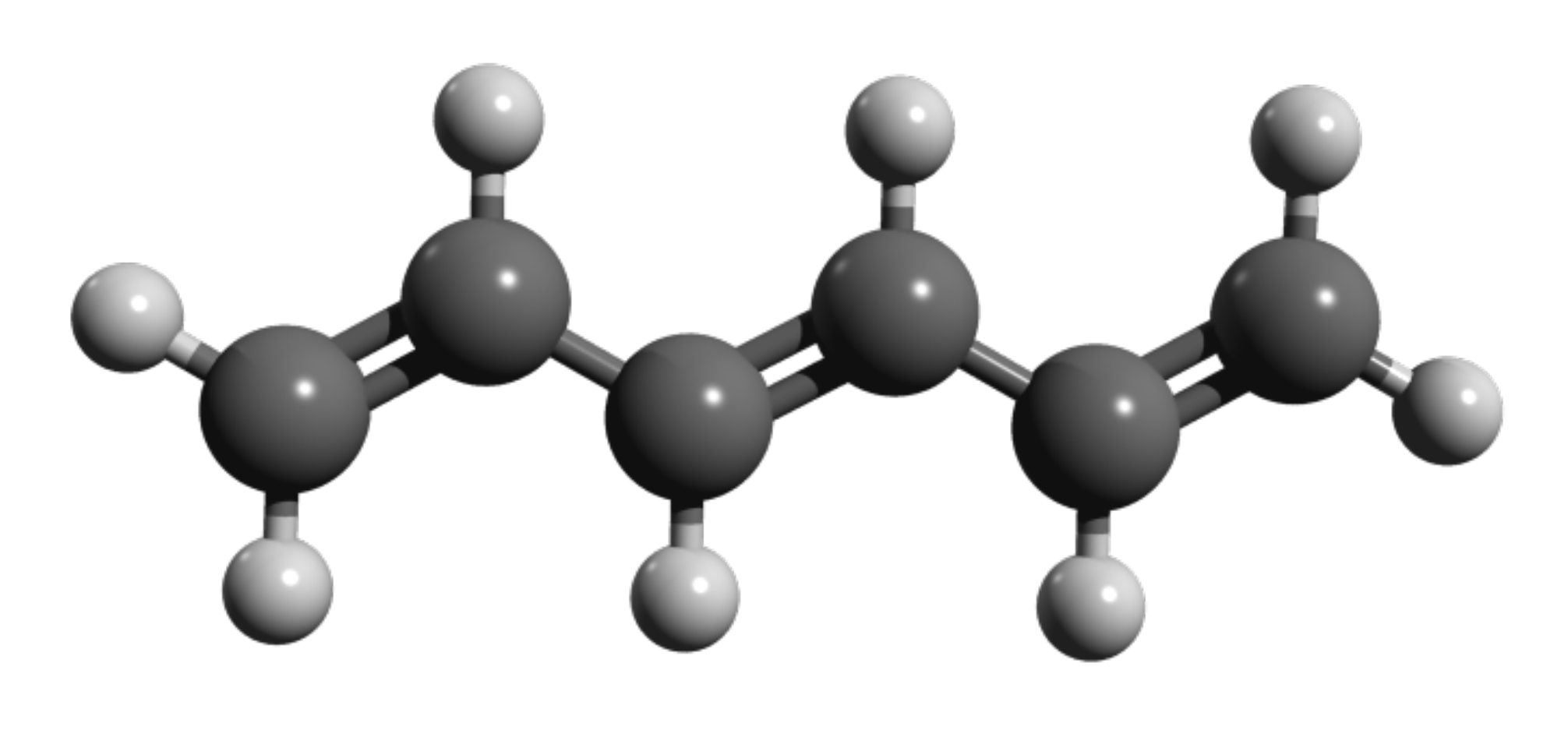} & \includegraphics[height=1cm]{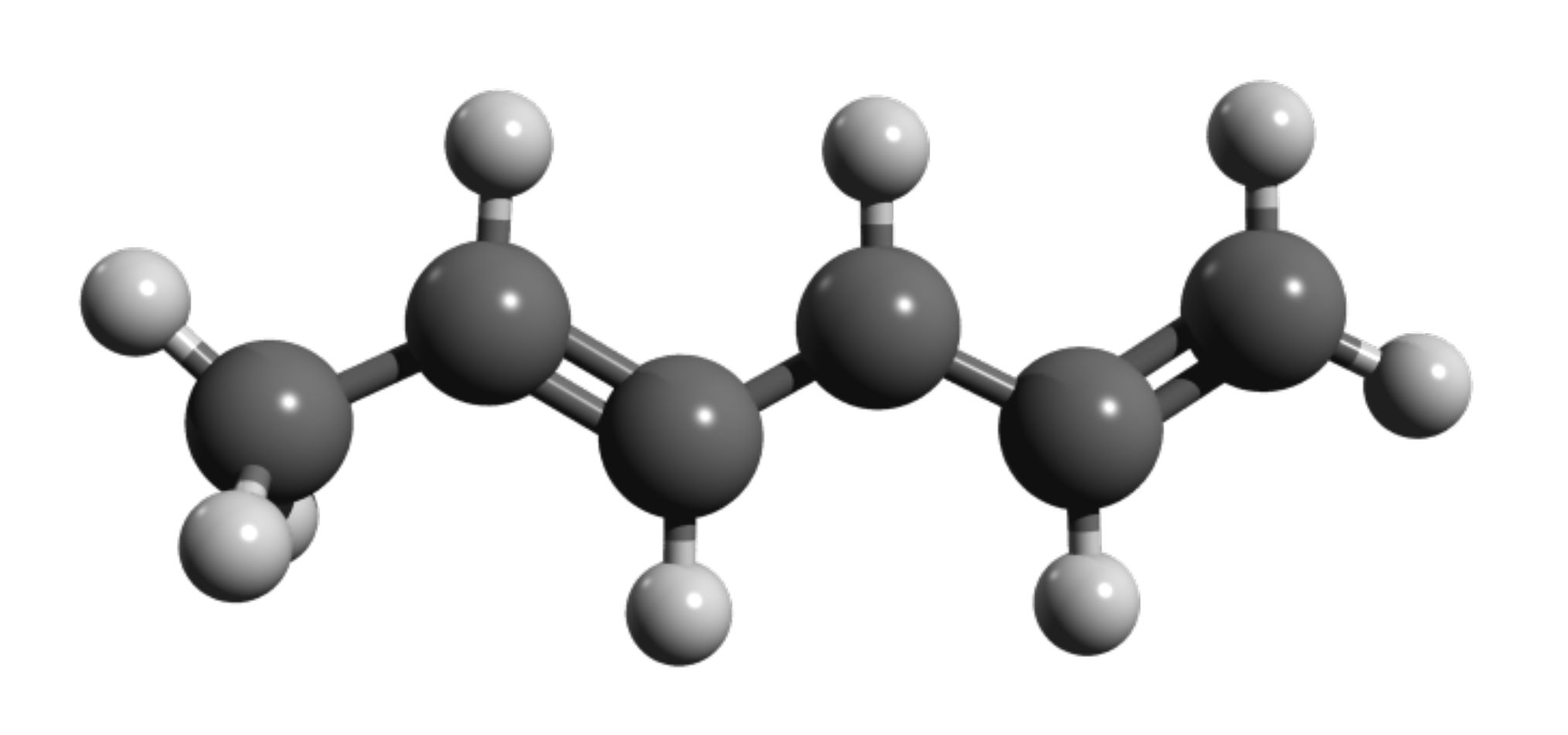} \\
 
  \end{tabular*}
\end{table*}

In the right panel of Figure \ref{fig:MSgrowth} we show a mass spectrum from our MD simulations of 3\,keV Ar atoms colliding with [C$_4$H$_6$]$_{50}$ clusters. Qualitatively the simulations agree with the experimental results in that there are many different growth products formed in the collisions. However, the simulations predict much broader distributions in the number of H atoms in products with any given number of C atoms than what we observe in the experiments (especially for the products with 7 or less C atoms). As was seen in experiments on clusters of pyrene \cite{Delaunay:2015aa}, the main growth mechanism is that individual molecules are promptly destroyed by the impinging projectile (knockout) and that the fragments rapidly form bonds with neighboring molecules on sub-picosecond timescales. The differences we see here between the experiments and simulations with 3\,keV Ar suggest that secondary fragmentation, where the newly formed molecules undergo additional fragmentation, plays a much more important role here than in the case of molecular growth inside pyrene clusters. The simulated and experimental results are much closer to each other in the case with pyrene clusters\cite{Delaunay:2015aa}. Another important factor for the survival of reaction products could be the role of charge, which is not included in the present classical simulations. For larger molecules, like PAHs and fullerenes with their many delocalized electrons, the reactivity and stability is largely the same regardless of whether the molecule is neutral or a singly-charged cation \cite{PhysRevA.89.062708}. This is, however, likely not the case for smaller molecules like those that we detect here. The presence of ``magic'' (particularly stable/abundant) growth products in the experimental mass spectrum (left panel of Figure \ref{fig:MSgrowth}) does indeed suggest that the charge of the molecules affects their stability, in particular since the strongest features (C$_4$H$_7^+$, C$_5$H$_7^+$, C$_6$H$_5^+$, C$_6$H$_7^+$, C$_6$H$_9^+$, C$_7$H$_7^+$, C$_7$H$_9^+$, C$_7$H$_{11}^+$ etc.) all have closed-shell electronic structures. Despite the differences between the experimental and theoretical results, both are consistent with prompt knockout of individual atoms acting as the main mechanism responsible for driving the first steps of the bond-forming reactions. However, we can not rule out that other processes, e.g.\ heating of the clusters though electronic scattering or through the formation of radicals by ionization, also play a role in the growth processes we observe. As an example, photoionization of ethylene (C$_2$H$_4$) clusters by synchrotron radiation has been shown to induce molecular growth in those systems\cite{Franceschi:2002aa}. 


The ``magic'' structures in our experimental results indicate that only the most stable molecule produced in the collisions survive long enough to be detected on the present experimental time scale of tens of microseconds. In order to determine the most stable structures for the growth products detected, we have performed Density Functional Theory (DFT) calculations of all possible isomers for a given number of C and H atoms. Due to the increasingly large number of isomers possible for some of the larger systems, e.g.\ C$_7$H$_6$ alone has 1230 unique isomers, we have limited our search to C$_5$H$_x^+$, where $x=6,7,8$, and C$_6$H$_x^+$, where $x=5,6,7,8,9$. In Table \ref{tbl:dft} we show the lowest energy open (linear or branched carbon backbone) and cyclic (with rings consisting of 5 or 6 C atoms) structures for these systems determined at the B3LYP/CC-pVDZ level of theory. For all of these systems, with the exception of C$_5$H$_8^+$, cyclic structures are more stable than the lowest energy open structures. We see that the energy difference between open and cyclic structures varies significantly, and is nearly 2\,eV in the case of C$_6$H$_6^+$. Our classical MD simulations show that most growth products are created when small fragments form covalent bonds with intact C$_4$H$_6$ molecules. Products formed in this way will initially have open structures, which in many cases have high potential energies relative to the cyclic ground states. As the reaction products relax into lower energy structures, this excess potential energy will be converted into vibrational energy that could lead to secondary fragmentation (together with the energy deposited in the collisions). This, together with the large energy differences between open and cyclic isomers could help explain why some products, e.g.\ C$_5$H$_6^+$ and C$_6$H$_6^+$, are less abundant than other similarly sized products.

\begin{figure}[h]
\centering
    \includegraphics[width=1\columnwidth]{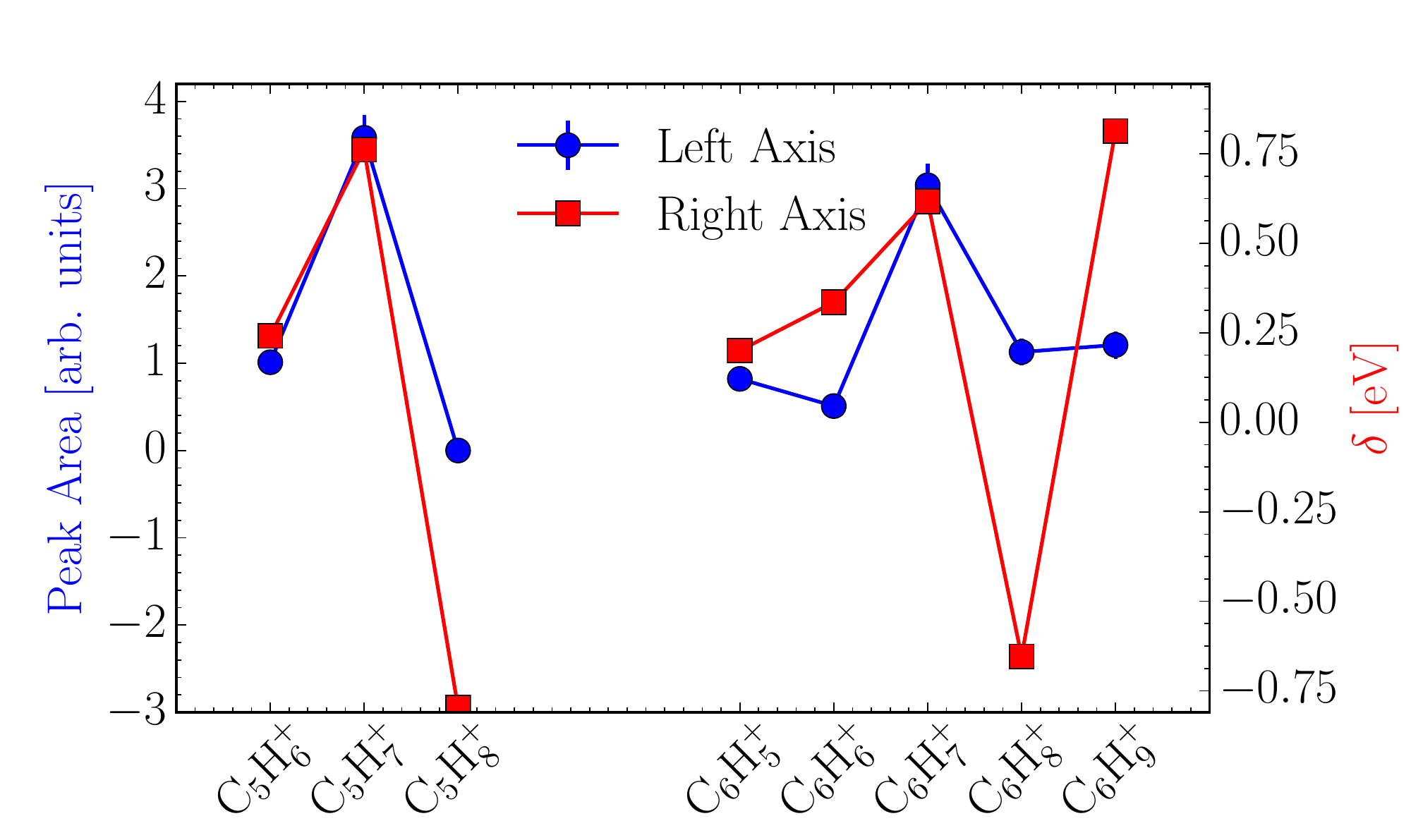}
  \caption{Left vertical scale: Integrated peak intensities (blue circles) of C$_5$H$_x^+$ and C$_6$H$_x^+$ reaction products following 3\,keV Ar$^+$ + [C$_4$H$_6$]$_k$ collisions. Right vertical scale: Relative stabilities of C$_5$H$_x^+$ and C$_6$H$_x^+$ as defined by Eq.\ \ref{eq:delta} in the text (red squares). 
  }
  \label{fig:areadetla}
\end{figure}

Using the energies from our DFT calculations, we follow Ref.\cite{Zettergren:2010aa}\ and define a parameter $\delta$ for each reaction product as
\begin{equation}
\delta = 0.5 \times \{E(\text{C}_n\text{H}_{x+1}^+) + E(\text{C}_n\text{H}_{x-1}^+)\} - E(\text{C}_n\text{H}_{x}^+),
\label{eq:delta}
\end{equation}
where $E(\text{C}_n\text{H}_{x}^+)$ is the total electronic energy of the most stable isomer of the C$_n$H$_{x}^+$ ion. This term allows us to compute the stability (in terms of the enthalpy of formation) of a molecule relative to the mean of its nearest neighbors in size, i.e.\ molecules with one less H atom and one additional H atom. Positive values of $\delta$ indicates that a molecular ion is (relatively) more stable than its neighbors and in the same way ions that are less stable than their neighbors will have negative values of $\delta$. In Figure \ref{fig:areadetla} we show the calculated $\delta$-values for reaction products with 5 and 6 C atoms, together with the integrated areas of the respective peaks in the experimental mass spectrum for the 3\,keV Ar$^+$ collisions (left panel of Fig.\ \ref{fig:MSgrowth}). In Figure \ref{fig:areadetla} we see that the oscillations in the $\delta$-values between different molecules agree with the odd-even effects that we find in the abundances of reaction products formed in the experiments. This qualitative agreement supports the idea that an initially broad size distribution of hot reaction products in our simulated mass spectrum can result in distributions with ``magic'' features (as seen in the experiments) through losses of H atoms. However, since $\delta$ only compares the stability of a molecule with those nearest in size, it essentially represents the relative stability of these molecules against the loss of single H atoms. It is also likely that the molecules formed in a decaying cluster could lose not only individual H atoms, but also one or more C atoms or H$_2$ molecules. This means that a complete theoretical description of the secondary fragmentation that reshapes the reaction products would have to consider other statistical fragmentation channels in addition to single H-loss (see e.g.\ Refs.\cite{Sanchez:2016aa,Aguirre:2017aa}).

%
%
%

\section{Summary and Conclusions}
We have shown that keV ions colliding with loosely bound clusters of 1,3-butadiene can induce bond-forming reactions. As in the case of ion-induced reactions in PAH\cite{Delaunay:2015aa} and fullerene clusters\cite{PhysRevLett.110.185501,:/content/aip/journal/jcp/139/3/10.1063/1.4812790}, the present reactions could also be initiated when atoms are knocked out of molecules via Rutherford-like ion-atom scattering processes. The fragments that are initially formed in such ways are often highly reactive and form covalent bonds with neighboring molecules on sub-picosecond timescales. Our classical MD simulations indicate that a wide range of molecular growth products are formed also in the present collisions between 3\,keV Ar$^+$ ions and clusters of butadiene (C$_4$H$_6$) molecules. However, in the experiment we mainly observe particularly stable, ``magic'' structures, which most likely is due to fast dissociation of the less stable reaction products.. The most abundant of the growth products which we observe are C$_{5}$H$_{7}^+$ and C$_{6}$H$_{7}^+$. Our DFT calculations show that the most stable isomers of these (and of other abundant reactions products) have ring structures. Further, the measured final size distribution of the reaction products can be understood in terms of their calculated relative stabilities. These findings suggest that we are indeed forming cyclic rings from initially linear molecules. The similarities between our experimental and simulated mass spectra suggest that the very fast atom knockout processes that are included in these simulations play important roles. However, it cannot be excluded that also other types of processes such as electronic excitation processes or ionization processes also can initiate molecular growth---in particular in clusters of smaller molecules.


The reactions that we have discussed in this work are likely to occur whenever energetic particles with suitable combinations of mass, charge, and kinetic energy collide with aggregates of matter, such as when supernova shockwaves interact with carbon-based dust grains in the interstellar medium (ISM)\cite{Micelotta:2010aa}. In this context it is important to note that argon (Ar) is significantly heavier than the ionic species most commonly found in space\cite{Micelotta:2010aa}. However, the present molecular growth products are believed to be quite general and will most likely occur for different types of projectiles depending on the amount of energy deposited through nuclear scattering. In general, this processes is the dominant mechanism for energy transfer in slow collisions (like those discussed here). We thus expect the knockout driven molecular growth to be important in shocks, where H and He atoms/ions with kinetic energies up to a few hundred eV, and C atoms/ions with energies up to a few keV are abundant\cite{Micelotta:2010aa}. In the future, it will be interesting to perform a systematic investigation of the relative importance of ionization and of electronic and nuclear stopping processes for molecular growth inside clusters of small hydrocarbon molecules. 

\section*{Acknowledgements}
This work was supported by the Swedish Research Council (Contract No.\ 2016-06625, No.\ 2016-04181, and No.\ 621-2015-04990). Research was conducted in the framework of the International Associated Laboratory (LIA) Fragmentation DYNAmics of complex MOlecular systems --- DYNAMO. We acknowledge the COST action CM1204 XUV/X-ray light and fast ions for ultrafast chemistry (XLIC). AM is funded by the European Union Seventh Framework Programme (PEOPLE2013-ITN-ARGENT project) under the grant agreement no.\ 608163. GDA is funded by the H2020 project ITN-EJD-642294 (TCCM: Theoretical Chemistry and Computational Modelling).




\bibliography{Library} 
\bibliographystyle{rsc} 

\end{document}